\documentclass[journal]{IEEEtran}
\ifCLASSINFOpdf
\else
\fi

{}
{}
\newtheorem{remark}{Remark}{}
{}
\newtheorem{proposition}{Proposition}{}
\usepackage{amsmath,amssymb,amsfonts,bm,mathtools,makecell,hyperref,cite,stfloats,multirow,subcaption,float}
\usepackage[ruled, lined, linesnumbered, commentsnumbered, longend]{algorithm2e}
\usepackage{xcolor,indentfirst,color}
\newcounter{MYtempeqncnt}
\makeatletter
\newcommand{\Rmnnum}[1]{\expandafter\@slowromancap\romannumeral#1@}
\newcommand\relphantom[1]{\mathrel{\phantom{#1}}}
\makeatother

\SetCommentSty{mycommfont}

\makeatletter
\renewcommand\normalsize{%
	\@setfontsize\normalsize\@xpt\@xiipt
	\abovedisplayskip 7\p@ \@plus2\p@ \@minus5\p@
	\abovedisplayshortskip \z@ \@plus3\p@
	\belowdisplayshortskip 6\p@ \@plus3\p@ \@minus3\p@
	\belowdisplayskip \abovedisplayskip
	\let\@listi\@listI}
\makeatother

\usepackage[belowskip=-1pt,aboveskip=0pt]{caption}
\begin{document}

%
\title{Distributed Extended Object Tracking Using Coupled Velocity Model from WLS Perspective}
%
%
%

\author{Zhifei~Li,~Yan~Liang, and ~Linfeng~Xu,~\IEEEmembership{Member,~IEEE}
	\thanks{Z. Li is with School of Space Information, Space Engineering University, Beijing 101416, China (e-mail: lee@seu.email.cn).}	
	\thanks{Y. Liang and L. Xu are with the Laboratory of Information Fusion
		Technology, School of Automation, Northwestern Polytechnical University,
		Xi’an 710072, China (e-mail: liangyan@nwpu.edu.cn; xulinf@gmail.com). }
	\thanks{Manuscript received November 23, 2021; revised March 28, 2022; accepted May 3, 2022. Date of publication May 20, 2022; date of current version June 2, 2022. This work was supported by the National Natural Science Foundation of China under Grants 61873205 and 61771399. The associate editor coordinating the review of this manuscript and approving it for publication was Dr.Wee Peng Tay. (Corresponding author: Zhifei Li.)}}

%
%

\markboth{IEEE TRANSACTIONS ON SIGNAL AND INFORMATION PROCESSING OVER NETWORKS,~VOL.~8, 2022}%
{Shell \MakeLowercase{\textit{et al.}}: Bare Demo of IEEEtran.cls for IEEE Journals}
%



\maketitle

\begin{abstract}

This study proposes a coupled velocity model (CVM) that establishes the relation between the orientation and velocity using their correlation, avoiding that the existing extended object tracking (EOT) models treat them as two independent quantities. As a result, CVM detects the mismatch between the prior dynamic model and actual motion pattern to correct the filtering gain, and simultaneously becomes a nonlinear and state-coupled model with multiplicative noise. The study considers CVM to design a feasible distributed weighted least squares (WLS) filter. The WLS criterion requires a linear state-space model containing only additive noise about the estimated state. To meet the requirement, we derive such two separate pseudo-linearized models by using the first-order Taylor series expansion. The separation is merely in form, and the estimates of interested states are embedded as parameters into each other's model, which implies that their interdependency is still preserved in the iterative operation of two linear filters. With the two models, we first propose a centralized WLS filter by converting the measurements from all nodes into a summation form. Then, a distributed consensus scheme, which directly performs an inner iteration on the priors across different nodes, is proposed to incorporate the cross-covariances between nodes. Under the consensus scheme, a distributed WLS filter over a realistic network with ``naive'' node is developed by proper weighting of the priors and measurements. Finally, the performance of proposed filters in terms of accuracy, robustness, and consistency is testified under different prior situations. 
\end{abstract}


%
\IEEEpeerreviewmaketitle

\section{Introduction}
%
%
%
%
\IEEEPARstart{W}{ith} increased resolution capabilities of modern sensors (e.g., phased array radar), multiple measurements from different scattering source of an object appear in a detection process \cite{b1}. In this situation, one can fuse these available measurements to get a joint estimate on the kinematic state (e.g., position and velocity) and extent (e.g., size and orientation) about the object. This induces a so-called extended object tracking (EOT) problem \cite{b2,b3}. The previous works on the EOT system rely on different state-space models, such as the random matrix (RM) model for elliptical extents \cite{b4,b5,RM21,zhang20,zhang21}, random hyper-surface model (RHM) \cite{b6} and Gaussian process (GP) model \cite{b7,b8,b9} for star-convex extents, and multiplicative error model (MEM) \cite{b10,b11} for axis symmetric extents, etc.  

In recent years, wireless sensor network (WSN) has received much attention as it uses a cooperative protocol to increase the perception capability from different field-of-view (FoV) \cite{b12,b13,b14,b15,survey12}. In general, WSN involves two types of network architectures, i.e., the centralized and distributed. To derive a centralized EOT filter under the RM model, G. Vivone et al. integrated multiple sensors’ measurements into the fusion center to output a fused estimate \cite{b16,b17,b18}. The centralized architecture provides an optimal estimate, while a key challenge is that the fusion center suffers from a computational burden for large-scale sensor networks. The other challenge is that the fusion service will be suspended or even denied if the fusion center does not work properly (e.g., under a network attack). 

The distributed architecture discards the fusion center, so it overcomes these challenges to some extent \cite{constrain11,non-cooperate16,maximum12}. In \cite{b19}, a distributed EOT filter was proposed by minimizing the weighted Kullback-Leibler divergence. To accomplish an asynchronous measurements fusion, Liu et al. proposed a distributed EOT particle filter, where the Gaussian mixture approximations of local posterior density functions were fused by a named geometric mean fusion rule \cite{b20}. Recently, Hua et al. derived a distributed variational Bayesian filter for the statistical characteristic identification and joint estimation \cite{b21}. Therein, the alternating direction method of multipliers was used to hold a consensus estimate. Apart from these filters under the RM model \cite{b19,b20,b21}, Ren et al. used the diffusion strategy to provide a distributed EOT filter within the MEM model \cite{bb}.    
 
The above mentioned distributed filters over sensor networks have the following common shortcomings. First, the models (i.e., the RM and MEM model) they rely on do not establish a tight relation between estimated states from the kinematic perspective, which causes the EOT merely being a joint estimation problem. In fact, an EO's extent including the orientation enables to point out whether its velocity has changed, since a change in the orientation must be caused by its velocity. Second, an EO is detected by each sensor node during the whole tracking process. This is definitely not suitable in a realistic scenario, since a sensor node has fixed FoV and limited sensing distance. Moreover, a network topology is usually sparse (not fully-connected), and thus the measurements are not available in a node and its immediate neighboring nodes (i.e., the node is a ``naive'' node). Hence, how to design a feasible distributed filter becomes very challenging, especially when combined with constrained communication resource. Third, the cross-covariances across different nodes are neglected as the computational cost and bandwidth requirements become unscalable for a large-scale network. Due to this reason, those existing distributed estimation schemes only yield a sub-optimal estimate. 

In this study, we endeavor to overcome the mentioned shortcomings and propose a distributed weighted least squares filter over a realistic sensor network with ``naive'' nodes. The main contributions are as follows. 
\begin{enumerate}
	\item We propose a state-space model named coupled velocity model (CVM). The CVM introduces a sideslip angle to integrate the orientation and velocity components so that their correlation is constructed via a more intuitive way. Using CVM will improve the performance of the related EOT filters, as it detects the mismatch between the actual motion pattern and prior dynamic model to correct the filtering gain. 
	\item By performing the first-order Taylor series expansion, we establish two separate pseudo-linearized measurement models with only additive noise. Compared with the original CVM model, the two models do not lose any first and second moment information, and the cross-correlation between estimated states is also preserved in each other's model. More importantly, they provide an efficient entry to derive the corresponding filters under the WLS criterion. 
	\item With the separate models, we derive a centralized WLS filter to simultaneously estimate the kinematic state and extent. To reduce the computational cost, the centralized filter converts the measurements from all of sensor nodes into a summation form.
	\item A consensus scheme is proposed to pave the way such that the cross-covariances across different nodes are sustained by directly performing an inner iteration on the priors. Under the scheme, we give a distributed WLS filter over a realistic network with ``naive'' nodes. To testify the robustness and effectiveness of the proposed filter, several numerical experiments are conducted under different scenarios.  
\end{enumerate}

For clarity, some notations that are used throughout the study are listed in Table \ref{tab}. The reminder of this study is organized as follows. Section \ref{sec:problem} gives a brief problem formulation. Section \ref{sec:sepe} presents two separate measurement models. Section \ref{sec:centralized} presents a centralized filter and Section \ref{sec:distributed}  presents the corresponding distributed filter. Numerical examples and results are presented in Section \ref{sec:simulation}. Section \ref{sec:conclusion} concludes this study.

\begin{table}[htpb]
	\renewcommand\arraystretch{1.5}
	\centering
	\setlength\tabcolsep{0.8pt}
	\caption{Notations}\label{tab}
	\begin{tabular}{c|c}
	\Xhline{1pt}
		Notation & Definition  \\ \hline
		$(\cdot)^{\mathsf{T}}$ & transpose of a matrix/vector \\ \hline
		$\Vert \cdot \Vert$ & $\mathit{l}^{2}$-norm \\ \hline
		$\mathbf I_n$ & $n$-th dimensional identity matrix\\ \hline
		$\mathrm{col}(\mathbf A_i)_{i \in \mathcal S}$ & stack $\mathbf A_i$ on top of each other to form a column matrix \\ \hline
		$\mathcal{N}$  & set of sensor nodes\\ \hline
		$\mathcal N^s \setminus \{s\}$ & set of neighboring nodes of the $s \in \mathcal{N}$ (excluding $s$ itself) \\ \hline
		$\bm x_k$  & kinematic state  \\ \hline
		$\bm p_k$ & extent vector \\ \hline
		$\Omega^{p}_k \; \Omega^{x}_k$ & information matrix  \\ \hline
		$\mathbf C^{p}_k\; \mathbf C^{x}_k$ & centralized covariance matrix \\	 \hline
		$\mathbf C_{k,ss}^{x(l)}\; \mathbf C_{k,ss}^{p(l)}$ &  covariance matrix on node $s$ at the $l$-th iteration  \\ \hline
		$\hat{\bm x}^{(l)}_{k,s} \; \hat{\bm p}^{(l)}_{k,s}$ & estimate on node $s$ at the $l$-th iteration \\  \hline
		$\hat{\bm x}_{k}\; \hat{\bm p}_{k}$ & centralized estimate \\  \hline 
		$\{\bm y_{k,s}^{i} \}^{n_{k,s}}_{i=1}$ & $n_k$ measurements on node $s$ \\  \hline
		$\mathcal Y_{k,s} $ & accumulated measurements on node $s $ \\  \hline
		$\mathcal Y_k$ & accumulated measurements from all sensor nodes \\
		\Xhline{1pt}
	\end{tabular}
\end{table}

\section{Problem Formulation}
\label{sec:problem}
\begin{figure}[htbp]
	\centering\includegraphics[width=0.45\textwidth]{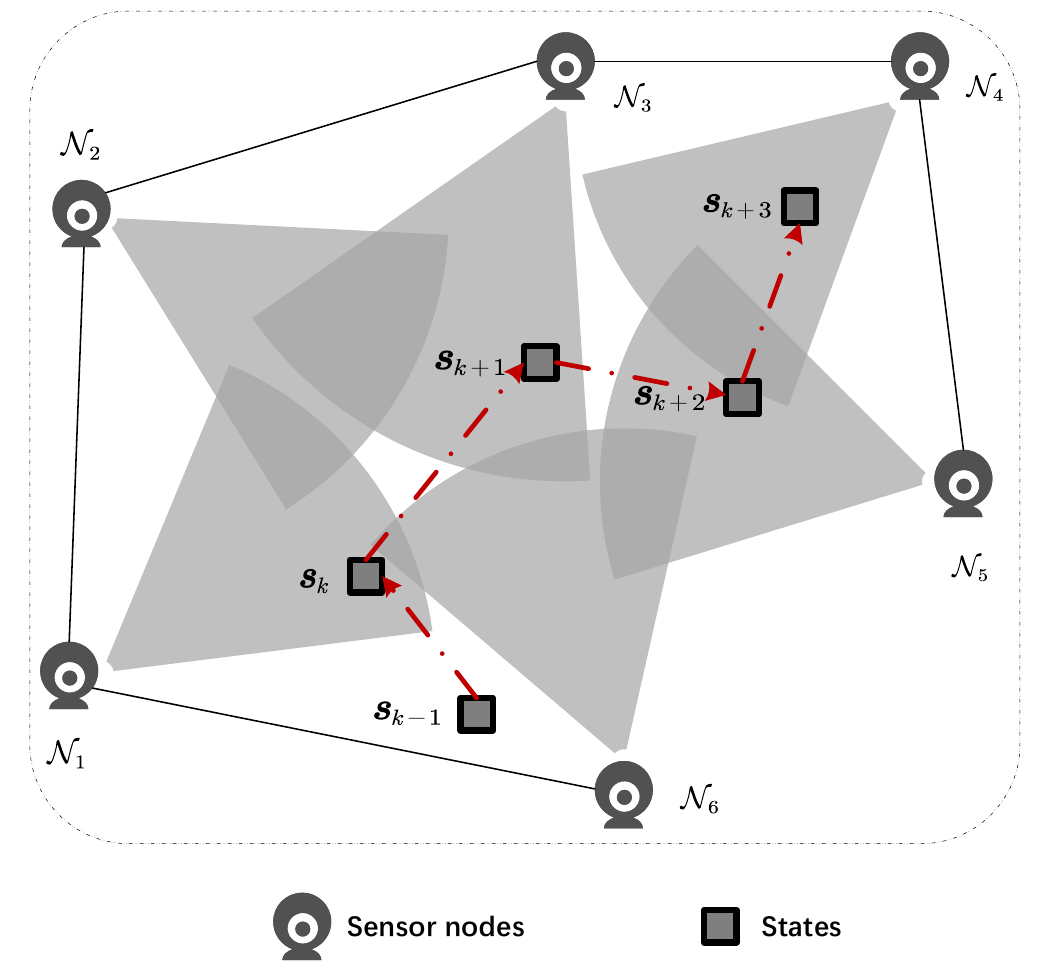}.
	\caption{\small{The figure shows a topology with $6$ sensor nodes $\{\mathcal N_1,\mathcal N_2,\dots,\mathcal N_6 \}$. The black solid lines represent the communication links between different nodes. Let the initial state of the object be $\bm s_{k-1}$, and then the object has four states $\{ \bm s_k,\bm s_{k+1},\bm s_{k+2}, \bm s_{k+3} \}$ over time. The figure also shows the existence of ``naive'' nodes. For example, when the object is in the state $\bm s_k$, only the node $\mathcal N_1$ observes the object. The nodes $\{\mathcal N_3,\mathcal N_4,\mathcal N_5 \}$ and their immediate neighboring nodes do not observe the object. Therefore, the nodes $\{\mathcal N_3,\mathcal N_4,\mathcal N_5 \}$ are the ``naive'' nodes for the state $\bm s_k$.}} 
	\label{fig1}
\end{figure}

 Here, the considered network topology $\mathcal{G}=(\mathcal{N, A})$ is shown in Fig. \ref{fig1}, where $\mathcal{N}$ is the set of sensor nodes, and $\mathcal{A \subseteq N \times N}$ is the set of edges such that $(s, j) \in \mathcal{A}$ if node $s$ communicates with $j$. For the node $s \in \mathcal{N}$, let $\mathcal N^s \setminus \{s\}$ be a set of its neighboring nodes (excluding $s$ itself). The accumulated sensor measurements on node $s \in \mathcal{N}$ at time $k$ are denoted as $\mathcal Y_{k,s} = \{\bm y_{k,s}^{i} \}^{n_{k,s}}_{i=1}$, and the accumulated measurements from all sensor nodes are denoted as $\mathcal Y_k = \{\mathcal Y_{k,s}\}_{s \in \mathcal{N}}$. 

Next, we propose a novel state-space model, namely coupled velocity model (CVM), where the correlation between the velocity and orientation is established  via a sideslip angle.\\
\noindent (1) {State Parameterization}

At time $k$, the kinematic state $\bm x_k$ 
\begin{equation} \label{eq1}
\bm x_k = [(\bm x_k^\mathrm{c})^{\mathsf{T}}, v_k^{\mathsf{x}}, v_k^{\mathsf{y}}, \cdots]^{\mathsf{T}}
\end{equation}
involves the centroid position $\bm x_k^\mathrm{c} \in \mathbb{R}^2$, the velocity $\bm [v_k^{\mathsf{x}},v_k^{\mathsf{y}}]^{\mathsf{T}} := \bm {\vartheta}_k$ in $x$ and $y$ axes, and possible quantities such as acceleration. As for the extent
\begin{equation} \label{eq2}
\bm p_k = [ l_{k,1}, l_{k,2},\beta_k]^{\mathsf{T}},
\end{equation} 
it involves the semi-lengths $l_{k,1}$ and $l_{k,2}$, and the sideslip angle $\beta_k = \arctan(\frac{v_k^{\mathsf{y}}}{v_k^{\mathsf{x}}}) - \alpha_k$ that represents a drift between the orientation $\alpha_k$ and velocity direction $ \arctan(\frac{v_k^{\mathsf{y}}}{v_k^{\mathsf{x}}})$.\\
\noindent (2) {Measurement Model}

At time $k$, the measurement $\bm y_{k,s}^{i}$ on node  $s\in\mathcal{N}$ is given in \eqref{eq3},
\begin{figure*}[!t]
		\normalsize
		\setcounter{MYtempeqncnt}{\value{equation}}	
		\setcounter{equation}{2}
\begin{equation} \label{eq3}
		\bm y_{k,s}^{i} = \mathbf{H} \bm x_k +  \underbrace{ \begin{bmatrix}
				\frac{v_k^{\mathsf{y}}\sin{\beta_k} + v_k^{\mathsf{x}}\cos{\beta_k}}{\Vert \bm {\vartheta}_k \Vert} & \frac{ \textcolor{red}{v_k^{\mathsf{x}} \sin{\beta_k} - v_k^{\mathsf{y}} \cos{\beta_k}}}{\Vert \bm {\vartheta}_k \Vert}  \\ \frac{v_k^{\mathsf{y}} \cos{\beta_k}-v_k^{\mathsf{x}} \sin{\beta_k}}{\Vert \bm {\vartheta}_k \Vert}   & \frac{v_k^{\mathsf{y}}\sin{\beta_k} + v_k^{\mathsf{x}}\cos{\beta_k}}{\Vert \bm {\vartheta}_k \Vert}  
			\end{bmatrix}
			\begin{bmatrix}
				l_{k,1} & 0 \\ 0 & l_{k,2}
		\end{bmatrix}}_{:= \mathbf S_k}
		\underbrace{{\begin{bmatrix}
					{h_{k,1}^{i}} \\{	h_{k,2}^{i}}
		\end{bmatrix}}}_{:= \bm h_{k}^{i}}  + \bm v_{k,s}^{i},
	\end{equation}
\setcounter{equation}{\value{MYtempeqncnt}}
\hrulefill
\vspace*{2pt}
\end{figure*}
\setcounter{equation}{3}
where the measurement matrix $\mathbf H = \left[ \mathbf I_2 \; \bm 0 \right]$ extracts the position component from $\bm x_k$, the coefficient matrix $\mathbf S_k$ compacts the kinematics and extent, and the multiplicative noise $\bm h_{k,s}^{i}$ enables any scattering source lying on the boundary or interior of the object. It is assumed that $\bm v_{k,1}^{i}, \bm v_{k,2}^{i}, \cdots$ are uncorrelated zero-mean Gaussian noises with covariances $\mathbf C_s^v \delta_{sj}$ if $s=j$, $\delta_{sj}=1$, and $\delta_{sj}=0$, otherwise. Fig. \ref{fig2} gives an illustration for the model \eqref{eq3}.
\begin{figure}[htpb]
	\centering\includegraphics[width=0.45\textwidth]{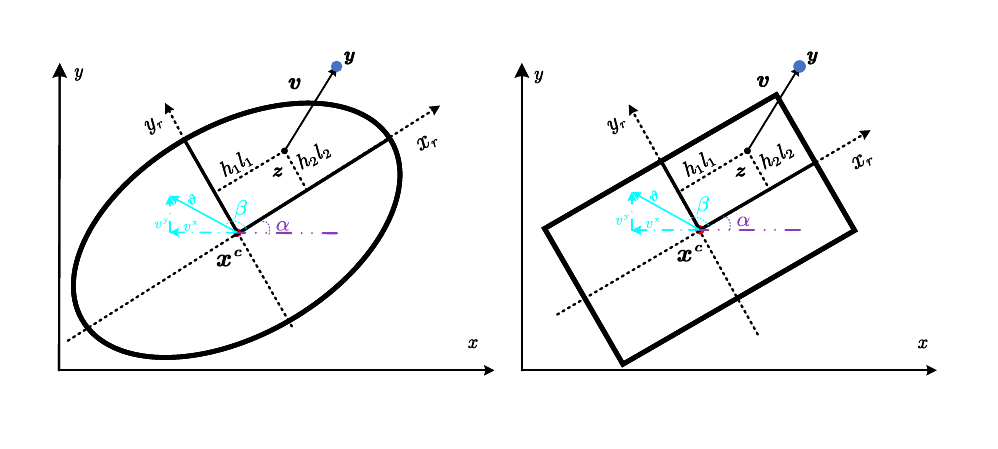}
	\caption{\small{Illustration of the measurement model. The time index $k$, measurement index ${i}$, and sensor node index $s$ are omitted in the figure. The centroid position of the object is $\bm x^{\mathrm{c}}$, and its extent and velocity are denoted as $\bm p = [ l_1,l_2,\beta]^{\mathsf{T}}$ and $\bm {\vartheta} = \bm [v^{\mathsf{x}},v^{\mathsf{y}}]^{\mathsf{T}}$, respectively. By counterclockwise rotating an angle $\alpha = \arctan ({v^{\mathsf{y}}}/{v^{\mathsf{x}}}) - \beta$ (i.e., the orientation) along $x$-axis, a reference coordinates $x_r$-$y_r$ is obtained. The scattering source $\bm z$ is compacted by the parameters $\bm p$, $\bm x^{\mathrm{c}}$ and multiplicative noise $\bm h = [h_1,h_2]^{\mathsf{T}}$. The measurement $\bm y$ is obtained by the source $\bm z$ plus a Gaussian measurement noise $\bm v$. The illustration gives a direct view about the correlation between the velocity and orientation. It can be seen that the model is feasible to describe the perpendicular axis-symmetric extents, such as an ellipse or a rectangle.}} 
	\label{fig2}
\end{figure}\\
\noindent (3) {Dynamic Models}

The dynamic models for the kinematics and extent are given as follows:
\begin{equation} \label{eq4}
\bm x_{k+1} = \mathbf {\Phi}_k^x \bm x_k + \bm w_k^x
\end{equation} 
\begin{equation} \label{eq5}
\bm p_{k+1} = \mathbf {\Phi}_k^p \bm p_k + \bm w_k^p
\end{equation}
where $\mathbf {\Phi}_k^x$ and $\mathbf {\Phi}_k^p$ are transition matrices, and $\bm w_k^x$ and $\bm w_k^p$ are zero-mean Gaussian process noises with covariances $\mathbf C_w^x$ and $\mathbf C_w^p$, respectively. One can select the corresponding transition matrices according to the actual motion pattern and body structure, e.g., for a rigid object with nearly constant velocity, 
\begin{equation*}
\mathbf {\Phi}_k^x = \begin{bmatrix}
1  & 0 & \mathrm{T} & 0 \\ 0 & 1 & 0 & \mathrm{T} \\ 0 & 0 & 1 & 0 \\ 0 & 0 & 0 & 1 
\end{bmatrix}, \qquad \mathbf {\Phi}_k^p = \mathbf I_3.
\end{equation*}
\begin{remark} \label{rem1}
	\begin{itemize}
		\item Compared with the previous model in \cite{b11}, \eqref{eq3} uses the sideslip angle $\beta_k$ to establish a tight relation between the velocity components $\bm [v_k^{\mathsf{x}}\;v_k^{\mathsf{y}}]$ and orientation $\alpha_k$ from the kinematic perspective. On the one hand, the relation contributes to finding the mismatch between the actual motion pattern and prior dynamic model. A main reason is that the orientation provides an intuitive insight to point out whether the motion pattern has changed. Once the pattern has changed, the corresponding filter should set more weight to the measurement instead of the prior. On the other hand, the relation is capable of describing the object drift cases such as in maritime radar applications \cite{LiuEM21,Liu21}.
		\item The dynamic model \eqref{eq5} w.r.t the extent involves the process noise $\bm w_k^p$. Although the study focuses on a rigid object (i.e., its extent is time-invariant), there may exist a distortion of extent to some extent in a sensor's FoV during the tracking process. Thus, the noise $\bm w_k^p$ is introduced to describe the distortion.
	\end{itemize}
\end{remark}
Although the state-space model is given, directly using the model to achieve a centralized or distributed filter still confronts some intractable difficulties. For the centralized filter, a main difficulty is how to process the massive measurements from multiple nodes especially in the EOT scenario. The distributed filter over a network with ``naive'' nodes faces two difficulties: 1) how to properly balance the weight between the priors and measurements to give a consensus estimate; 2) how to preserve the cross-covariances among the nodes in a distributed tracking system, especially for a large-scale network, so that the distributed estimate closes to the corresponding centralized value (i.e., the optimal estimate) as much as possible. 

\section{Separation of measurement model with coupled kinematics and extent}
\label{sec:sepe}
This study introduces the WLS criterion to handle those aforementioned difficulties. On the one hand, the criterion facilitates the centralized filter to reduce its computational cost by converting the massive data into a compact summation form. On the other hand, it allows the local estimate on each node to be given a proper weight in the final result. However, it requires a linear state-space model with only additive noise about the estimated state. Thus, we separate \eqref{eq3} into two pseudo-linearized models to satisfy the requirement.

Here, the measurements $\{\bm y_{k,s}^{i} \}^{n_{k,s}}_{i=1}$ on node $s \in \mathcal N $ are processed sequentially. Let $ \hat{\bm x}_{k}^{[i-1]}$, $\hat{\bm p}_{k}^{[i-1]}$, $\mathbf C_{k}^{x[i-1]}$ and $\mathbf C_{k}^{p[i-1]}$ denote the prior estimates for the kinematics $\bm x_{k}$ and extent $\bm p_{k}$ plus their corresponding covariances at the ${[i-1]}$-th sequential operation. The node $s$  processes $\bm y_{k,s}^{i}$ to obtain the updated estimates $ \hat{\bm x}_{k}^{[i]}$, $\hat{\bm p}_{k}^{[i]}$, $\mathbf C_{k}^{x[i]}$ and $\mathbf C_{k}^{p[i]}$. Next, we focus on presenting two pseudo-linearized measurement models w.r.t $\bm x_k$ and $\bm p_k$, respectively.

\begin{proposition}[Separate model I] \label{the1}
	The measurement model about $\bm x_k$ is 
	\begin{equation} \label{eq6}
		\bm y_{k,s}^{i} \approx \mathbf{H} \bm x_k + \bm v_{k,s}^{x[i]},
	\end{equation}
where $\bm v_{k}^{x[i]}$ is the equivalent noise with $\mathbb{E} (\bm v_{k,s}^{x[i]}) = \bm 0 $, $\mathrm{Cov} (\bm v_{k,s}^{x[i]}) = \mathbf R_{k,s}^{x[i]} := \mathbf C^{\mathrm{\Rmnnum{1}}} + \mathbf C^{\mathrm{\Rmnnum{2}}} + \mathbf C^{\mathrm{\Rmnnum{3}}} + \mathbf C_s^v$. The terms $\mathbf C^{\mathrm{\Rmnnum{1}}}$, $\mathbf C^{\mathrm{\Rmnnum{2}}}$, and $\mathbf C^{\mathrm{\Rmnnum{3}}}$ are given as follows:
	\begin{equation} \label{eq7}
	\mathbf C^{\mathrm{\Rmnnum{1}}} = \hat{\mathbf S}_k^{[i-1]} \mathbf C^h \left(\hat{\mathbf S}_k^{[i-1]}\right)^{\mathsf{T}},
\end{equation}

\begin{equation} \label{eq8}
	\underbrace{[\epsilon_{mn}]}_{\mathbf C^{\mathrm{\Rmnnum{2}}}} = \mathrm{tr} \left\{ \mathbf C_k^{p[i-1]} \left(\hat{\mathbf J}_{nk,p}^{[i-1]}\right)^{\mathsf{T}}\mathbf C^h \hat{\mathbf J}_{mk,p}^{[i-1]}\right\},
\end{equation}

\begin{equation} \label{eq9}
	\underbrace{[\varpi_{mn}]}_{\mathbf C^{\mathrm{\Rmnnum{3}}}} = \mathrm{tr} \left\{\mathbf H \mathbf C_k^{x[i-1]} \mathbf H^{\mathsf{T}} \left(\hat{\mathbf J}_{nk,v}^{[i-1]}\right)^{\mathsf{T}}\mathbf C^h \hat{\mathbf J}_{mk,v}^{[i-1]}\right\},
\end{equation}
for $m,n \in \{1,2\}$. The intermediate quantities $\hat{\mathbf J}_{1k,p}^{[i-1]}$, $\hat{\mathbf J}_{2k,p}^{[i-1]}$, $\hat{\mathbf J}_{1k,v}^{[i-1]}$, and $\hat{\mathbf J}_{2k,v}^{[i-1]}$ are the Jacobian matrices of the first row $\mathbf S_{1,k}$ and second row $\mathbf S_{2,k}$ of $\mathbf S_k$ around the ${[i-1]}$-th extent estimate $\hat{\bm p}_k^{[i-1]}$ and velocity estimate $\hat{\bm {\vartheta}}_k^{[i-1]}$, respectively.
\end{proposition}
\begin{IEEEproof}
	See Appendix \ref{proof1}.
\end{IEEEproof}

Notice that the quantities $\mathbf C^{\mathrm{\Rmnnum{1}}}$, $\mathbf C^{\mathrm{\Rmnnum{2}}}$, and $\mathbf C^{\mathrm{\Rmnnum{3}}}$ in \eqref {eq6} are treated as constant terms at the $[i]$-th sequential operation since they are calculated based on the former estimates $\hat{\bm p}_k^{[i-1]}$ and $\hat{\bm {\vartheta}}_k^{[i-1]}$.

Due to the existence of zero-mean multiplicative noise $\bm h_{k}^{i}$, a pseudo-measurement using 2-fold Kronecker product is required to update the extent \cite{b11,b22}. The ${[i]}$-th pseudo-measurement $\mathbf Y_k^{[i]}$ is given as
\begin{equation} \label{eq10}
	\mathbf Y_{k,s}^{[i]} = \mathbf F \left( (\bm y_{k,s}^{i} - \mathbf H \hat{\bm x}_k^{[i-1]})\otimes (\bm y_{k,s}^{i} - \mathbf H \hat{\bm x}_k^{[i-1]}) \right),
\end{equation}
and its expectation is
\begin{equation} \label{eq11}
\mathbb{E} (\mathbf Y_{k,s}^{[i]})	=  \mathbf F \mathrm{vect} \left(\mathbf C_{k,s}^{y[i]}\right),
\end{equation} 
residual covariance is 
\begin{equation} \label{eq12}
	\mathrm{Cov} (\mathbf Y_{k,s}^{[i]}) = \mathbf F \left( \mathbf C_{k,s}^{y[i]}\otimes \mathbf C_{k,s}^{y[i]} \right) (\mathbf F + \tilde{\mathbf F})^{\mathsf{T}}
\end{equation}
with
\begin{equation} \label{eq13}
	\mathbf F = \begin{bmatrix}
		1  & 0 & 0 & 0 \\ 0 & 0 & 0 & 1 \\ 0 & 1 & 0 & 0
	\end{bmatrix}, \quad 	\tilde{\mathbf F} = \begin{bmatrix}
		1 & 0 & 0 & 0 \\ 0 & 0 & 0 & 1 \\ 0 & 0 & 1 & 0
	\end{bmatrix}.
\end{equation}

 \begin{proposition}[Separate model II] \label{the2}
 	 According to \eqref{eq10}, the measurement model about $\bm p_k$ is
	\begin{equation} \label{eq14}
 		\mathbf Y_{k,s}^{[i]} \approx \hat{\mathbf M}_k^{[i-1]} \bm p_k + \bm v_{k,s}^{p[i]}
 	\end{equation}
 	where 
 	\begin{equation} \label{eq15}
	\hat{\mathbf M}_k^{[i-1]} = \begin{bmatrix}
		2 \hat{\mathbf S}_{1,k}^{[i-1]} \mathbf C^h \hat{\mathbf J}_{1k,p}^{[i-1]} \\
		2 \hat{\mathbf S}_{2,k}^{[i-1]} \mathbf C^h \hat{\mathbf J}_{2k,p}^{[i-1]} \\
		\hat{\mathbf S}_{1,k}^{[i-1]} \mathbf C^h \hat{\mathbf J}_{2k,p}^{[i-1]} +
		\hat{\mathbf S}_{2,k}^{[i-1]} \mathbf C^h \hat{\mathbf J}_{1k,p}^{[i-1]}
	\end{bmatrix},
	 \end{equation}
 and $\bm v_{k,s}^{p[i]}$ is the equivalent noise with
 	\begin{equation} \label{eq16}
 		\mathbb E (\bm v_{k,s}^{p[i]})=\bar{\bm v}_{k,s}^{p[i]}:=\mathbf F \mathrm{vect} \left(\mathbf C_{k,s}^{y[i]}\right) - \hat{\mathbf M}_k^{[i-1]} \hat{\bm p}_k^{[i-1]},
 	\end{equation}
 	\begin{equation} \label{eq17}
 		\begin{split}
 		\mathrm{Cov} (\bm v_{k,s}^{p[i]}) = \mathbf R_{k,s}^{p[i]} := & \mathbf F \left( \mathbf C_{k,s}^{y[i]}\otimes \mathbf C_{k,s}^{y[i]} \right) (\mathbf F + \tilde{\mathbf F})^{\mathsf{T}} \\
 		& - \hat{\mathbf M}_k^{[i-1]} \mathbf C_k^{p[i-1]} \hat{\mathbf M}_k^{[i-1]{\mathsf{T}}}.
 		\end{split}
 	\end{equation}
 \end{proposition}
\begin{IEEEproof}
	See Appendix \ref{proof2}.
\end{IEEEproof}	
 
\begin{remark}
	\begin{itemize}
 		\item Although the models \eqref{eq6} and \eqref{eq14} are separated from \eqref{eq3}, the interdependency between the kinematics $\bm x_k$ and extent $\bm p_k$ still remains as parameters in each other's model. In this condition, the joint estimation of $\bm x_k$ and $\bm p_k$ is transferred into an iterative operation of related linear filters.
		\item Considering the complex calculation in the higher-order terms, such as the Hessian matrix, the models \eqref{eq6} and \eqref{eq14} only utilize the first-order Taylor series expansion in \eqref{eq47}. Moreover, the model \eqref{eq14} merely depends on \eqref{eq47} as an essential prerequisite, otherwise \eqref{eq14} cannot be obtained.
	\end{itemize}
\end{remark}
\section{Centralized WLS Filter}
\label{sec:centralized}    
This section presents a centralized weighted least squares filter (CWLSF) as a benchmark for the following distributed filter. The CWLSF first integrates the models \eqref{eq6} and \eqref{eq14} into the WLS structure, and then the estimated states are alternately updated in two linear filters.
\subsection{Measurement Update}
In the measurement update, the accumulated measurements $\mathcal Y_k = \{\mathcal Y_{k,s}\}_{s \in \mathcal{N}}$ from all nodes are gathered into the fusion center. For clarity, assume that there are $n_k$ measurements at time $k$ for each node $s \in \mathcal N$. Given the $[i-1]$-th prior estimates, the fusion center sequentially processes $\{\bm y_{k,s}^{i}\}_{s \in \mathcal{N}}$ to give the updated estimates. According to \eqref {eq6}, define the central measurement, measurement matrix, noise covariance, and noise information matrix related to the measurement set $\{\bm y_{k,s}^{i}\}_{s \in \mathcal{N}}$ as
\begin{equation}\label{eq18}
	\left \{ \begin{lgathered} 
		\bm y_{k,\mathrm{c}}^{i} := \mathrm {col} (\bm y_{k,1}^{i},\bm y_{k,2}^{i},\cdots,\bm y_{k,\lvert \mathcal N \rvert}^{i});\;
		\mathbf H_{\mathrm{c}} := [\mathbf H;\mathbf H;\cdots;\mathbf H] \\
		\bm {\nu}_{k,\mathrm{c}}^{x {[i]}} := \mathrm {col} (\bm {v}_{k,1}^{x {[i]}},\bm {\nu}_{k,2}^{x {[i]}},\cdots,\bm {v}_{k,\lvert \mathcal N \rvert}^{x {[i]}})\\
		\mathbf R_{k,\mathrm{c}}^{x[i]} := \mathrm {diag} (\mathbf R_{k,1}^{x[i]},\mathbf R_{k,2}^{x[i]},\cdots,\mathbf R_{k,\lvert \mathcal N \rvert}^{x[i]}) \\
		\mathbf V_{k,\mathrm{c}}^{x[i]} := \mathrm{diag}(\mathbf V_{k,1}^{x[i]},\dots,\mathbf V_{k,\lvert \mathcal N \rvert}^{x[i]})= (\mathbf R_{k,\mathrm{c}}^{x[i]})^{-1}
	\end{lgathered} \right. 
\end{equation}
where the subscript ``c" denotes ``central", and $\lvert \mathcal N \rvert$ is the cardinality of $\mathcal N$. If a node $s$ does not obtain measurements, let $\bm y_{k,s}^{i} = \bm 0$, $\mathbf V_{k,s}^{x[i]} = \bm 0$.

According to \eqref {eq14}, define the central measurement, measurement matrix, noise covariance, and noise information matrix related to the pseudo-measurement set $\{{\mathbf Y}_{k,s}^{[i]}\}_{s \in \mathcal{N}}$ as
\begin{equation}\label{eq19}
	\left \{ \begin{lgathered} 
		\tilde{\mathbf Y}_{k,s}^{[i]} := \mathbf Y_{k,s}^{[i]} - \bar{\bm {v}}_{k,s}^{p {[i]}} ,\: \tilde{\bm {v}}_{k,s}^{p {[i]}} := \bm {v}_{k,s}^{p {[i]}} - \bar{\bm {v}}_{k,s}^{p {[i]}}\\
		\tilde{\mathbf Y}_{k,\mathrm{c}}^{[i]} := \mathrm {col} (\tilde{\mathbf Y}_{k,1}^{[i]},\tilde{\mathbf Y}_{k,2}^{[i]},\cdots,\tilde{\mathbf Y}_{k,\lvert \mathcal N \rvert}^{[i]})\\
		\mathbf M_{k,\mathrm{c}}^{[i-1]} := [\hat{\mathbf M}_{k}^{[i-1]};\hat{\mathbf M}_{k}^{[i-1]};\cdots;\hat{\mathbf M}_{k}^{[i-1]}]\\
		\bm {\nu}_{k,\mathrm{c}}^{p {[i]}} := \mathrm {col} (\tilde{\bm {v}}_{k,1}^{p {[i]}},\tilde{\bm {v}}_{k,2}^{p {[i]}},\cdots,\tilde{\bm {v}}_{k,\lvert \mathcal N \rvert}^{p {[i]}}) \\
	    \mathbf R_{k,\mathrm{c}}^{p[i]} = \mathrm {diag} (\mathbf R_{k,1}^{p[i]},\mathbf R_{k,2}^{p[i]},\cdots,\mathbf R_{k,\lvert \mathcal N \rvert}^{p[i]})  \\
		\mathbf V_{k,\mathrm{c}}^{p[i]} := \mathrm{diag}(\mathbf V_{k,1}^{p[i]},\ldots,\mathbf V_{k,\lvert \mathcal N \rvert}^{p[i]})= (\mathbf R_{k,\mathrm{c}}^{p[i]})^{-1}
	\end{lgathered} \right.. 
\end{equation}
If a node $s$ does not obtain measurements, let $\tilde{\mathbf Y}_{k,s}^{[i]} = \bm 0$, $\mathbf V_{k,s}^{p[i]} = \bm 0$.

Define
\begin{itemize}
	\item the prior error of kinematics ${\bm x}_k$ as $\tilde{\bm x}_k^{[i-1]}:=\hat{\bm x}^{[i-1]}_{k} - {\bm x}_k$,
	\item the corresponding covariance as $\mathrm{Cov} (\tilde{\bm x}_k^{[i-1]}) := \mathbf{C}^{x[i-1]}_{k}$,
	\item the information matrix as $\mathbf{\Omega}^{x[i-1]}_{k} := \left(\mathbf C_{k}^{x[i-1]}\right)^{-1}$.
\end{itemize}

Similarly, define 
\begin{itemize}
	\item the prior error of extent ${\bm p}_k$ as $\tilde{\bm p}_k^{[i-1]}:=\hat{\bm p}^{[i-1]}_{k} - {\bm p}_k$,
	\item  the corresponding covariance as $\mathrm{Cov} (\tilde{\bm p}_k^{[i-1]}) := \mathbf{C}^{p[i-1]}_{k}$,
	\item the information matrix as $\mathbf{\Omega}^{p[i-1]}_{k} := \left(\mathbf C_{k}^{p[i-1]}\right)^{-1}$.
\end{itemize}
Here, the notation $(\cdot)_{k}^{[0]}$ denotes the predicted estimate at time $k$.

Combining \eqref{eq18} with prior error $\tilde{\bm x}_k^{[i-1]}$ yields 
\begin{equation} \label{eq20}
	\begin{bmatrix}
		\hat{\bm x}^{[i-1]}_{k} \\ \bm y_{k,\mathrm{c}}^{i}
	\end{bmatrix} =
	\begin{bmatrix}
		\mathbf I \\ \mathbf H_{\mathrm{c}}
	\end{bmatrix} \bm x_k + 
	\begin{bmatrix}
		\tilde{\bm x}_k^{[i-1]} \\ \bm {\nu}_{k,c}^{x {[i]}}
	\end{bmatrix}.
\end{equation}

The centralized WLS estimate of the kinematics $\bm x_k$ plus information matrix are
\begin{subequations} \label{eq21}
	\begin{equation} \label{eq21a}
		\begin{split}
		\hat{\bm x}^{[i]}_{k} = & \left( \mathbf{\Omega}^{x[i-1]}_{k} +  \sum_{s \in \mathcal N} \mathbf H^{\mathsf{T}} \mathbf V_{k,s}^{x[i]} \mathbf H \right)^{-1} \\
		 & \times \left( \mathbf{\Omega}^{x[i-1]}_{k} \hat{\bm x}^{[i-1]}_{k} +
		\sum_{s \in \mathcal N} \mathbf H^{\mathsf{T}} \mathbf V_{k,s}^{x[i]} \bm y_{k,s}^{i} \right) ,
		\end{split}
	\end{equation}
	\begin{equation} \label{eq21b}
		\mathbf{\Omega}^{x[i]}_{k} = \mathbf{\Omega}^{x[i-1]}_{k} + \sum_{s \in \mathcal N} \mathbf H^{\mathsf{T}} \mathbf V_{k,s}^{x[i]} \mathbf H.
	\end{equation}
\end{subequations}
Also, combining \eqref{eq19} with prior error  $\tilde{\bm p}_k^{[i-1]}$ yields
\begin{equation} \label{eq22}
	\begin{bmatrix}
		\hat{\bm p}^{[i-1]}_{k} \\ \tilde{\mathbf Y}_{k,\mathrm{c}}^{[i]}
	\end{bmatrix} =
	\begin{bmatrix}
		\mathbf I \\ \mathbf M_{k,\mathrm{c}}^{[i-1]}
	\end{bmatrix}\bm p_k + 
	\begin{bmatrix} 
		\tilde{\bm p}_k^{[i-1]} \\ \bm {\nu}_{k,c}^{p {[i]}}
	\end{bmatrix}.
\end{equation}
The centralized WLS estimate of the extent $\bm p_k$ plus information matrix are
\begin{subequations} \label{eq23}
	\begin{equation} \label{eq23a}
		\begin{split}
		\hat{\bm p}^{[i]}_{k} = & \left( \mathbf{\Omega}^{p[i-1]}_{k} + \sum_{s \in \mathcal N}  \left( \hat{\mathbf M}_{k}^{[i-1]}\right) ^{\mathsf{T}} \mathbf V_{k,s}^{p[i]} \hat{\mathbf M}_{k}^{[i-1]} \right)^{-1} \\
		& \times \left( \mathbf{\Omega}^{p[i-1]}_{k} \hat{\bm p}^{[i-1]}_{k} +
		\left( \hat{\mathbf M}_{k}^{[i-1]}\right) ^{\mathsf{T}} \mathbf V_{k,s}^{p[i]} \tilde{\mathbf Y}_{k,s}^{[i]} \right) ,
		\end{split}
	\end{equation}
	\begin{equation} \label{eq23b}
		\mathbf{\Omega}^{p[i]}_{k} = \mathbf{\Omega}^{p[i-1]}_{k} + \sum_{s \in \mathcal N}  \left( \hat{\mathbf M}_{k}^{[i-1]}\right) ^{\mathsf{T}} \mathbf V_{k,s}^{p[i]} \hat{\mathbf M}_{k}^{[i-1]}.
	\end{equation}
\end{subequations}
Notice that the measurements from all nodes are processed via a summation form as shown in \eqref{eq21a} and \eqref{eq23a}, which overcomes the data congestion in the centralized system to some extent.
\subsection{Time Update}
\begin{figure}[htpb]
	\centering
	\includegraphics[width=3.5in,height=2.1in]{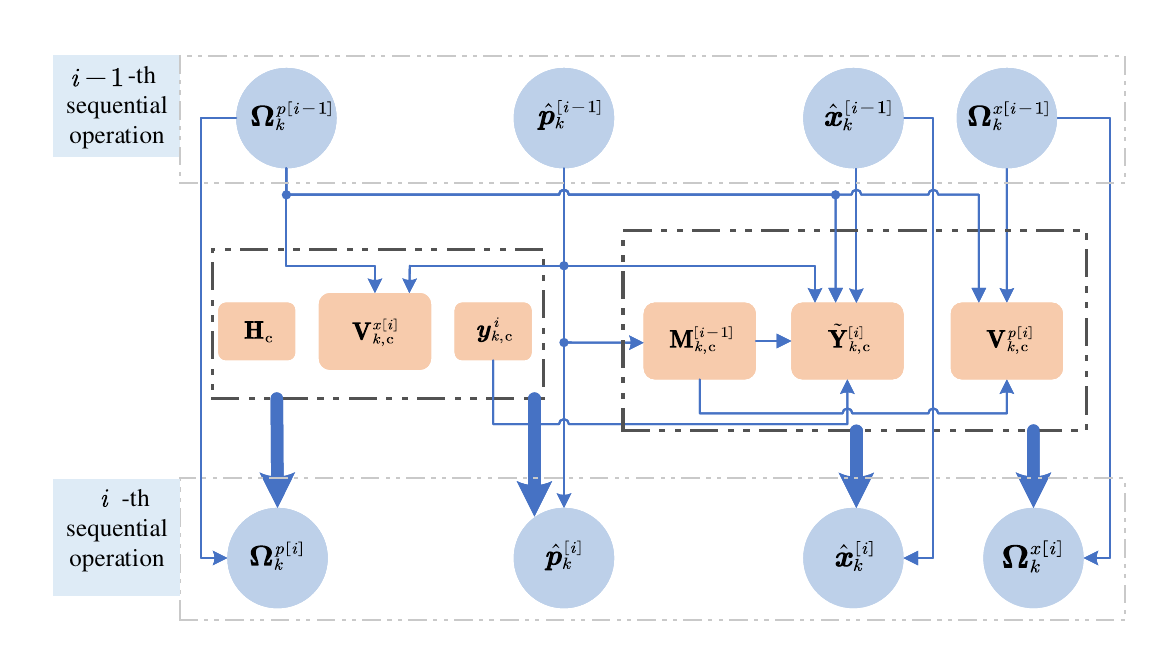}.
	\caption{\small{An illustration of the interdependency between $\bm x_k$ and $\bm p_k$.}}
	\label{sequen}
\end{figure}

After the $n_k$ batch of measurements are processed sequentially, \eqref{eq24} and \eqref{eq25} are performed to accomplish the time update. Since the temporal evolutions of both the kinematics and extent follow a linear model, the predicted estimates have the same form as in the information filter \cite{b24}, i.e., 
\begin{subequations} \label{eq24}
	\begin{equation} \label{eq24a}
		\hat{\bm x}^{[0]}_{k+1} = \mathbf {\Phi}_k^x \hat{\bm x}^{[n_k]}_{k} ,
	\end{equation}
	\begin{equation} \label{eq24b}
		\mathbf{\Omega}_{k+1}^{x[0]} = \left( \mathbf {\Phi}_k^x \left(\mathbf{\Omega}_{k}^{x[n_k]}\right)^{-1} \left(\mathbf {\Phi}_k^x \right)^{\mathsf{T}} + \mathbf C_w^x \right)^{-1} ,
	\end{equation}
\end{subequations}
\begin{subequations} \label{eq25}
	\begin{equation} \label{eq25a}
		\hat{\bm p}^{[0]}_{k+1} = \mathbf {\Phi}_k^p \hat{\bm p}^{[n_k]}_{k} ,
	\end{equation}
	\begin{equation} \label{eq25b}
		\mathbf{\Omega}_{k+1}^{p[0]} = \left( \mathbf {\Phi}_k^p \left(\mathbf{\Omega}_{k}^{p[n_k]}\right)^{-1} \left(\mathbf {\Phi}_k^p \right)^{\mathsf{T}} + \mathbf C_w^p \right)^{-1} . 
	\end{equation}
\end{subequations}
\begin{algorithm}[t]
	\caption{CWLSF Filter}\label{algorithm1}	
	\SetAlgoLined
	\textbf{Initialization:} $\hat{\bm x}_{1}^{[0]}$, $\hat{\bm p}_{1}^{[0]}$, $\mathbf{\Omega}_{1}^{x[0]}$, and $\mathbf{\Omega}_{1}^{p[0]}$ \;
	\For{$k \gets 1,2,\cdots $ \tcp*[h]{scan time} }{
	\textbf{Data:} $\{\bm y_{k,s}^{i} \}^{n_{k}}_{i=1} (s \in \mathcal N)$ \;
	Initialization: $\hat{\bm x}_{k}^{[0]}$, $\hat{\bm p}_{k}^{[0]}$, $\mathbf{\Omega}_{k}^{x[0]}$, and $\mathbf{\Omega}_{k}^{p[0]}$ \;
	
	\For{$i=1,2,\cdots,n_k$ \tcp*[h]{sequential}}{compute $\hat{\bm x}_{k}^{[i]}, \mathbf{\Omega}_{k}^{x[i]}, \hat{\bm p}_{k}^{[i]}, \mathbf{\Omega}_{k}^{p[i]}$ via \eqref{eq21} and \eqref{eq23}}
	\textbf{Output1:} $\hat{\bm{x}}_k \gets \hat{\bm x}_k^{[n_k]}, \mathbf C_{k}^{x} \gets \left( \mathbf{\Omega}_{k}^{x[n_k]}\right)^{-1}$ \;
	\textbf{Output2:} $\hat{\bm{p}}_k \gets \hat{\bm p}_k^{[n_k]},\mathbf C_{k}^{p} \gets \left( \mathbf{\Omega}_{k}^{p[n_k]}\right)^{-1}$ \;
	compute $\hat{\bm x}_{k+1}^{[0]},\mathbf{\Omega}_{k+1}^{x[0]},\hat{\bm p}_{k+1}^{[0]},\mathbf{\Omega}_{k+1}^{p[0]}$ via \eqref{eq24} and \eqref{eq25}	
	}
\end{algorithm}
 The detailed CWLSF is collected in Algorithm \ref{algorithm1}. It is important to note, that using \eqref{eq6} and \eqref{eq14} to achieve the corresponding filters poses that the interdependency between $\bm x_k$ and $\bm p_k$ exists in the $[i]$-th and $[i-1]$-th sequential operation. Here, Fig. \ref{sequen} gives an illustration of the interdependency in CWLSF.   

\section{Distributed WLS Filter}
\label{sec:distributed}
For a distributed system, a primary goal is to make the estimates between nodes achieve consensus. Meanwhile, the estimate on each node converges to the corresponding centralized value. However, due to the naivety and resource constraints, the goal may not be fully achieved at any given time.

Here, we design such a distributed WLS filter (DWLSF). Therein, DWLSF directly operates a consensus on the prior estimate set $\{\hat{\bm x}^{[i-1]}_{k,s},\hat{\bm p}^{[i-1]}_{k,s}\}_{s \in \mathcal{N}}$ to retain the cross-covariances across all the nodes, and then the updated estimate set $\{\hat{\bm x}^{[i]}_{k,s},\hat{\bm p}^{[i]}_{k,s}\}_{s \in \mathcal{N}}$ yields an optimal estimates (if the priors are converged to the same value). Next, we just take $\hat{\bm x}^{[i-1]}_{k,s}$ for an example to discuss how to achieve this. 

\textit{Average Consensus} (AC) is a popular consensus scheme to compute the arithmetic mean $\frac{1}{\lvert \mathcal N \rvert}\sum_{s \in \mathcal{N}} a_s$ of a variable set $\{a_s\}_{s\in\mathcal N}$ \cite{b25}. Define the initial value on each node $s\in\mathcal N$ be $a_s(0)$. At the $l$-th iteration, each node $s$ updates its value using the following protocol
\begin{equation} \label{eq26}
		a_s(l) = a_s(l-1) + \xi \sum_{j\in\mathcal N^s \setminus \{s\}} \left( a_j(l-1) -a_s(l-1)\right).
\end{equation}
By iteratively doing this, the values at all the nodes converge to the average value $\frac{1}{\lvert \mathcal N \rvert}\sum_{s \in \mathcal{N}} a_s$. The rate parameter $\xi$ is chosen between $0$ and $\frac{1}{\triangle\max}$, where $\triangle\max$ is the maximum degree of the network $\mathcal{G}$.

Suppose that consensus is performed directly on the priors, at the $l$-th iteration, the estimate on node $s$ is described as follows (here, the superscript $[i-1]$ is omitted):
\begin{equation} \label{eq27}
	\hat{\bm x}^{(l)}_{k,s} = \underbrace{\bm x_k + \tilde{\bm x}_k}_{\hat{\bm x}_k} + \bm{\delta}_{k,s}^{x(l)}
\end{equation}
where $\bm{\delta}_{k,s}^{x(l)}$ denotes the bias between the local estimate $\hat{\bm x}^{(l)}_{k,s}$ and centralized estimate $\hat{\bm x}_k$, and consensus makes $\bm{\delta}_{k,s}^{x(l)} \to 0$ if the iteration $L$ approaches to $\infty$. 

The error covariance in \eqref{eq27} is defined as $\mathbf C_{k,ss}^{x(l)} := \mathbb{E} [ \bm{\eta}_{k,s}^{x(l)}  (\bm{\eta}_{k,s}^{x(l)})^{\mathsf{T}} ]$, where the prior error $ \bm{\eta}_{k,s}^{x(l)} = \tilde{\bm x}_k + \bm{\delta}_{k,s}^{x(l)}$. Similarly, the error cross-covariance for any pair of nodes $\{s,j\}$ is $\mathbf C_{k,sj}^{x(l)} := \mathbb{E} [\bm{\eta}_{k,s}^{x(l)}  (\bm{\eta}_{k,j}^{x(l)})^{\mathsf{T}}]$. We drop the iteration index $(l)$, and define the collective prior estimate, prior error, and coefficient matrix as 
\begin{equation}\label{eq28}
	\left \{ \begin{lgathered}
		\bm{\mathcal{X}}_{k,\mathrm{c}}^{[i-1]} := \left[ \left(\hat{\bm x}_{k,1}^{[i-1]}\right)^{\mathsf{T}}, \left(\hat{\bm x}_{k,2}^{[i-1]}\right)^{\mathsf{T}}, \cdots ,\left(\hat{\bm x}_{k,\vert\mathcal{N}\vert}^{[i-1]}\right)^{\mathsf{T}}  \right]^{\mathsf{T}}, \\
		\bm{\eta}_{k,\mathrm{c}}^{x[i-1]} := \left[\left(\bm{\eta}_{k,1}^{x[i-1]} \right)^{\mathsf{T}}, \left(\bm{\eta}_{k,2}^{x[i-1]} \right)^{\mathsf{T}}, \cdots, \left(\bm{\eta}_{k,\vert \mathcal N \vert}^{x[i-1]} \right)^{\mathsf{T}}\right]^{\mathsf{T}},\\
		\bm{\mathcal H}_I := \underbrace{\left[ \mathbf I, \mathbf I,\cdots,\mathbf I \right]^{\mathsf{T}}}_{\vert\mathcal{N}\vert }  . 
	\end{lgathered} \right.
\end{equation}
Combining \eqref{eq18} with \eqref{eq28} yields 
\begin{equation} \label{eq29}
	\begin{bmatrix}
		\bm{\mathcal{X}}_{k,\mathrm{c}}^{[i-1]} \\ \bm y_{k,\mathrm{c}}^{i}
	\end{bmatrix} =
	\begin{bmatrix}
		\bm{\mathcal H}_I \\ \mathbf H_{\mathrm{c}}
	\end{bmatrix} \bm x_k + 
	\begin{bmatrix}
		{\bm \eta}_{k,\mathrm{c}}^{x[i-1]} \\ \bm {\nu}_{k,\mathrm{c}}^{x {[i]}}
	\end{bmatrix}.
\end{equation}
Denote the covariance and information matrix of error ${\bm \eta}_{k,\mathrm{c}}^{x[i-1]}$, respectively, as follows

\begin{equation} \label{eq30}
	\mathbf{C}^{x[i-1]}_{k,\mathrm{c}} = \begin{bmatrix}
		\mathbf{C}^{x[i-1]}_{k,11} & \mathbf{P}^{x[i-1]}_{k,12} & \dots & \mathbf{C}^{x[i-1]}_{k,1 \vert \mathcal N \vert} \\
		\mathbf{C}^{x[i-1]}_{k,21} & \mathbf{P}^{x[i-1]}_{k,22} & & \vdots \\
		\vdots & & \ddots & \\
		\mathbf{C}^{x[i-1]}_{k,\vert \mathcal N \vert 1} & \dots & & \mathbf{C}^{x[i-1]}_{k,\vert \mathcal N \vert \vert \mathcal N \vert}
	\end{bmatrix},
\end{equation}
\begin{equation} \label{eq31}
	\mathbf{F}^{x[i-1]}_{k,\mathrm{c}} = \begin{bmatrix}
		\mathbf{F}^{x[i-1]}_{k,11} & \mathbf{F}^{x[i-1]}_{k,12} & \dots & \mathbf{F}^{x[i-1]}_{k,1 \vert \mathcal N \vert} \\
		\mathbf{F}^{x[i-1]}_{k,21} & \mathbf{F}^{x[i-1]}_{k,22} & & \vdots \\
		\vdots & & \ddots & \\
		\mathbf{F}^{x[i-1]}_{k,\vert \mathcal N \vert 1} & \dots & & \mathbf{F}^{x[i-1]}_{k,\vert \mathcal N \vert \vert \mathcal N \vert}
	\end{bmatrix}.
\end{equation}
From \eqref{eq29} and \eqref{eq31}, the centralized WLS estimate of the kinematics $\bm x_k$ and information matrix are given as 
\begin{equation} \label{eq32}
	\begin{split}
		\hat{\bm x}^{[i]}_{k} = &{} \left( \bm{\mathcal H}_I^{\mathsf{T}} \mathbf{F}^{x[i-1]}_{k,\mathrm{c}} \bm{\mathcal H}_I + \left( \mathbf H_{\mathrm{c}} \right)^{\mathsf{T}} \mathbf V_{k,\mathrm{c}}^{x[i]} \mathbf H_{\mathrm{c}} \right)^{-1} \\
		&\relphantom{=} {} \times \left( \bm{\mathcal H}_I^{\mathsf{T}} \mathbf{F}^{x[i-1]}_{k,\mathrm{c}} \bm{\mathcal X}_{k,\mathrm{c}}^{[i-1]} +
		\left( \mathbf H_{\mathrm{c}}\right) ^{\mathsf{T}} \mathbf V_{k,\mathrm{c}}^{x[i]} \bm y_{k,\mathrm{c}}^{i} \right) \\
		= &{} \left( \sum_{s \in \mathcal N} (\mathbf{F}^{x[i-1]}_{k,s} + \mathbf{U}^{x[i]}_{k,s})\right)^{-1} \! \sum_{s \in \mathcal N} (\mathbf{F}^{x[i-1]}_{k,s} \hat{\bm x}_{k,s}^{[i-1]} + \bm{u}^{x[i]}_{k,s}),
	\end{split} 
\end{equation}

\begin{equation} \label{eq33}
	\mathbf{\Omega}^{x[i]}_{k} = \sum_{s \in \mathcal N} \left( \mathbf{F}^{x[i-1]}_{k,s} + \mathbf{U}^{x[i]}_{k,s}\right) ,
\end{equation}
where 
\begin{equation}\label{eq34}
	\left \{ \begin{lgathered}
		\mathbf{F}^{x[i-1]}_{k,s} = \sum_{j \in \mathcal N} \mathbf{F}^{x[i-1]}_{k,js}, \quad
		\mathbf{U}^{x[i ]}_{k,s} = \mathbf H^{\mathsf{T}} \mathbf V_{k,s}^{x[i]} \mathbf H,\\ 
		\bm{u}^{x[i]}_{k,s} = \mathbf H^{\mathsf{T}} \mathbf V_{k,s}^{x[i]} \bm y_{k,s}^i.
	\end{lgathered} \right. 
\end{equation}

In an analogous definitions about the kinematics in \eqref{eq28} to \eqref{eq31}, the centralized WLS estimate of the extent $\bm p_k$ and information matrix are given as
\begin{equation} \label{eq35}
	\hat{\bm p}^{[i]}_{k} = \left( \sum_{s \in \mathcal N} (\mathbf{F}^{p[i-1]}_{k,s} + \mathbf{U}^{p[i]}_{k,s})\right)^{-1}  \sum_{s \in \mathcal N} (\mathbf{F}^{p[i-1]}_{k,s} \hat{\bm p}_{k,s}^{[i-1]} + \bm{u}^{p[i]}_{k,s}),
\end{equation}
\begin{equation} \label{eq36}
	\mathbf{\Omega}^{p[i]}_{k} = \sum_{s \in \mathcal N} \left( \mathbf{F}^{p[i-1]}_{k,s} + \mathbf{U}^{p[i]}_{k,s}\right) ,
\end{equation}
where 
\begin{equation}\label{eq37}
	\left \{ \begin{lgathered}
		\mathbf{F}^{p[i-1]}_{k,s} = \sum_{j \in \mathcal N} \mathbf{F}^{p[i-1]}_{k,js} , \quad
		\mathbf{U}^{p[i]}_{k,s} = \left(\hat{\mathbf M}_{k}^{[i-1]}\right)^{\mathsf{T}} \mathbf V_{k,s}^{p[i]} \hat{\mathbf M}_{k}^{[i-1]},\\ 
		\bm{u}^{p[i]}_{k,s} = \left( \hat{\mathbf M}_{k}^{[i-1]} \right)^{\mathsf{T}} \mathbf V_{k,s}^{p[i]} \tilde{\mathbf Y}_{k,s}^{[i]}.
	\end{lgathered} \right. 
\end{equation}

Notice that the centralized estimates fully account for the entire error covariances (i.e., the error covariances for all nodes and error cross-covariances between nodes), but which are unknown in a distributed system. In the following, we show how \eqref{eq32}, \eqref{eq33}, \eqref{eq35}, and \eqref{eq36} can be computed in a distributed way.

\subsection{Implementation in a distributed way }
To compute the summation terms in Eqns. (32-33,35-36) via AC operation, let us define 
\begin{gather} \label{eq38}
	\delta \mathbf{\Omega}^{x[i]}_{k,s} := \mathbf{F}^{x[i-1]}_{k,s} + \mathbf{U}^{x[i]}_{k,s} ,\;\delta \hat{\bm x}^{[i]}_{k,s} := \mathbf{F}^{x[i-1]}_{k,s} \hat{\bm x}_{k,s}^{[i-1]} + \bm{u}^{x[i]}_{k,s}, \\ \label{eq39} 
	\delta \mathbf{\Omega}^{p[i]}_{k,s} := \mathbf{F}^{p[i-1]}_{k,s} + \mathbf{U}^{p[i]}_{k,s},\;\delta \hat{\bm p}^{[i]}_{k,s} := \mathbf{F}^{p[i-1]}_{k,s} \hat{\bm p}_{k,s}^{[i-1]} + \bm{u}^{p[i]}_{k,s}. 
\end{gather}
If a node $s$ does not obtain measurements, let $\bm{u}^{x[i]}_{k,s} = \bm 0$, $\mathbf{U}^{x[i]}_{k,s} = \bm 0$, $\bm{u}^{p[i]}_{k,s} = \bm 0$, and $\mathbf{U}^{p[i]}_{k,s} = \bm 0$.

Performing $L$ iterations on AC operation ($L$ is given a priori), the estimate and information covariance of the kinematics $\bm x_k$ on node $s$ are 
\begin{subequations} \label{eq40}
	\begin{equation} \label{eq40a}
		\hat{\bm x}^{[i]}_{k,s} = \left(\delta \mathbf{\Omega}^{x[i]}_{k,s} (L) \right)^{-1} \delta \hat{\bm x}^{[i]}_{k,s} (L),
	\end{equation}
	\begin{equation} \label{eq40b}
		\mathbf{\Omega}^{x[i]}_{k,s} = \omega_{k,s}^{[i]}  \left(\delta \mathbf{\Omega}^{x[i]}_{k,s} (L) \right)^{-1},
	\end{equation}
\end{subequations}
 where $\omega_{k,s}^{[i]}$ is a suitable scalar weight to match the corresponding centralized estimate. When $L$ approaches to infinity, a reasonable choice is $\omega_{k,s}^{[i]} = \vert \mathcal N  \vert$. However, the choice may have some drawbacks if only a finite number of  iterations is performed. With this respect, the scalar weight is referred to \cite[eq.~(4)]{b26}.

Similarly, the estimate and information covariance of the extent $\bm p_k$ on  node $s$ are  
\begin{subequations} \label{eq41}
	\begin{equation} \label{eq41a}
		\hat{\bm p}^{[i]}_{k,s} = \left(\delta \mathbf{\Omega}^{p[i]}_{k,s} (L) \right)^{-1} \delta \hat{\bm p}^{[i]}_{k,s} (L),
	\end{equation}
	\begin{equation} \label{eq41b}
		\mathbf{\Omega}^{p[i]}_{k,s} = \omega_{k,s}^{[i]} \left(\delta \mathbf{\Omega}^{p[i]}_{k,s} (L) \right)^{-1}.
	\end{equation}
\end{subequations} 

The distributed estimate $\hat{\bm x}^{[i]}_{k,s} (\hat{\bm p}^{[i]}_{k,s})$ converges to the corresponding centralized value $\hat{\bm x}^{[i]}_{k} (\hat{\bm p}^{[i]}_{k})$ only if the covariance $\mathbf{F}^{x[i-1]}_{k,\mathrm{s}} (\mathbf{F}^{p[i-1]}_{k,\mathrm{s}})$ is known. However, computing $\mathbf{F}^{x[i-1]}_{k,\mathrm{s}}$ at each time on each node is unrealistic since it needs the knowledge of the entire covariance matrix $\mathbf{C}^{x[i-1]}_{k,\mathrm{c}}$ (see \eqref{eq30}). The following Proposition \ref{the3} provides a solution about how to get the cross-covariances among the nodes. 

\begin{proposition}[Distributed Consensus Scheme] \label{the3}
	If the iteration $L \to \infty$, the estimate on node $s$ converges to the centralized estimate, i.e., $\hat{\bm x}^{(l)}_{k,s} \rightarrow \hat{\bm x}_k$.
	By doing such an iteration between nodes, the cross-covariances among the nodes are retained when reaching consensus.
\end{proposition}
\begin{IEEEproof}
	According to the variables defined in \eqref{eq27}, consensus forces $\bm{\delta}_{k,s}^{x(l)} \rightarrow 0$ if the iteration $L \to \infty$, which guarantees $\hat{\bm x}^{(l)}_{k,s} \rightarrow \hat{\bm x}_k$. Meanwhile, as $L \rightarrow \infty$, the error covariance on node $s$ also converges to the centralized error covariance, i.e., $\mathbf C_{k,ss}^{x(l)} \rightarrow  \mathbf C^x_k$. By alternatively doing so, for any pair of nodes $\{s,j\}$, $\hat{\bm x}^{(l)}_{k,s}$ and $\hat{\bm x}^{(l)}_{k,j}$ becomes correlated as $\hat{\bm x}^{(l)}_{k,s} \rightarrow \hat{\bm x}^{(l)}_{k,j}$ causes the cross-covariance $\mathbf C_{k,sj}^{x(l)} \neq 0$. In short, $\mathbf C_{k,sj}^{x(l)} \rightarrow  \mathbf C^x_k$ if the iteration $L \rightarrow \infty$.
\end{IEEEproof}

Next, the core objective is to compute $\mathbf{F}^{x[i-1]}_{k,s}$ and $\mathbf{F}^{p[i-1]}_{k,s}$ based on the Proposition \ref{the3}. 
\subsection{Computations of $\mathbf{F}^{x[i-1]}_{k,s}$ and $\mathbf{F}^{p[i-1]}_{k,s}$}
Here, we take $\mathbf{F}^{x[i-1]}_{k,s}$ as an example to show how to compute it using only a node’s own prior information covariance in two special cases. The first case considers the converged priors, as the prior estimates on all nodes ultimately converge to the same value after a large-enough number of consensus iterations. The second case is under a condition where the prior estimates across the nodes are uncorrelated to each other. The condition is possible during the early scan times when the nodes have limited knowledge about the considered object, and thus they are initialized to random values. 
\subsubsection{Case I: Converged Priors} 
In Case I, the prior estimate on each node has converged to the centralized estimate at the previous scan time $k-1$. Thus, at time $k$, the prior on each node is the same and equals to the centralized value (i.e., $L$ is enough large so that $\bm{\delta}_{k,s}^{x(l)}$ in \eqref{eq27} equals to $0$ for all $s$). 

From \eqref{eq21b} and \eqref{eq33}, we get 
\begin{equation} \label{eq42}
	\sum_{s \in \mathcal N} \mathbf{F}^{x[i-1]}_{k,s} = \mathbf{\Omega}^{x[i-1]}_{k} = \sum_{s \in \mathcal N} \frac{\mathbf{\Omega}^{x[i-1]}_{k}}{\vert \mathcal N \vert}.
\end{equation}
Then, for the converged priors, we have $\mathbf{\Omega}^{x[i-1]}_{k} = \mathbf{\Omega}^{x[i-1]}_{k,s}$ for all $s$. Using this in \eqref{eq42}, $\mathbf{F}^{x[i-1]}_{k,s}$ is calculated as 
\begin{equation} \label{eq43}
	\mathbf{F}^{x[i-1]}_{k,s} =  \frac{\mathbf{\Omega}^{x[i-1]}_{k,s}}{\vert \mathcal N \vert}.
\end{equation}   
\subsubsection{Case II: Uncorrelated Priors}
If the priors between nodes are uncorrelated with each other, \eqref{eq31} reduces to a block diagonal matrix, i.e., $\mathbf{F}^{x[i-1]}_{k,\mathrm{c}} = \mathrm{blkdiag} (\mathbf{F}^{x[i-1]}_{k,11}, \mathbf{F}^{x[i-1]}_{k,22},\cdots,\mathbf{F}^{x[i-1]}_{k,\vert \mathcal N \vert \vert \mathcal N \vert})$. With the definition about $\mathbf{F}^{x[i-1]}_{k,s}$ in \eqref{eq34}, we have
\begin{equation} \label{eq44}
	\mathbf{F}^{x[i-1]}_{k,s} = \mathbf{\Omega}^{x[i-1]}_{k,s}.
\end{equation}
Similarly, \eqref{eq45a} and \eqref{eq45b} give how to compute $\mathbf{F}^{p[i-1]}_{k,s}$, respectively, under the Case I and Case II,
\begin{subequations} \label{eq45}
	\begin{equation} \label{eq45a}
		\mathbf{F}^{p[i-1]}_{k,s} =  \frac{\mathbf{\Omega}^{p[i-1]}_{k,s}}{\vert \mathcal N \vert},
	\end{equation}
	\begin{equation} \label{eq45b}
		\mathbf{F}^{p[i-1]}_{k,s} = \mathbf{\Omega}^{p[i-1]}_{k,s}.
	\end{equation}
\end{subequations}

It is worth noting that DWLSE yields a convergent result due to the following three reasons. First, AC operation provides a global average values when the number of iterations is sufficiently large \cite{b27}. Second, the proposed consensus scheme ensures that the priors on each node converge to the centralized estimate at the next scan time. And these convergent priors will further assist DWLSE to yield a convergent estimate at the subsequent time steps. Third, in essence, the terms $\delta \hat{\bm x}^{[i]}_{k,s}, \delta \mathbf{\Omega}^{x[i]}_{k,s}, \delta \hat{\bm p}^{[i]}_{k,s}, \delta \mathbf{\Omega}^{p[i]}_{k,s}$ exchanged between nodes are information matrix and information vector (information matrix multiplied by estimate). This corresponds to use the information matrix to set a suitable weight on the corresponding node.

\begin{figure*}[t]
	\centering
	\includegraphics[scale=0.4]{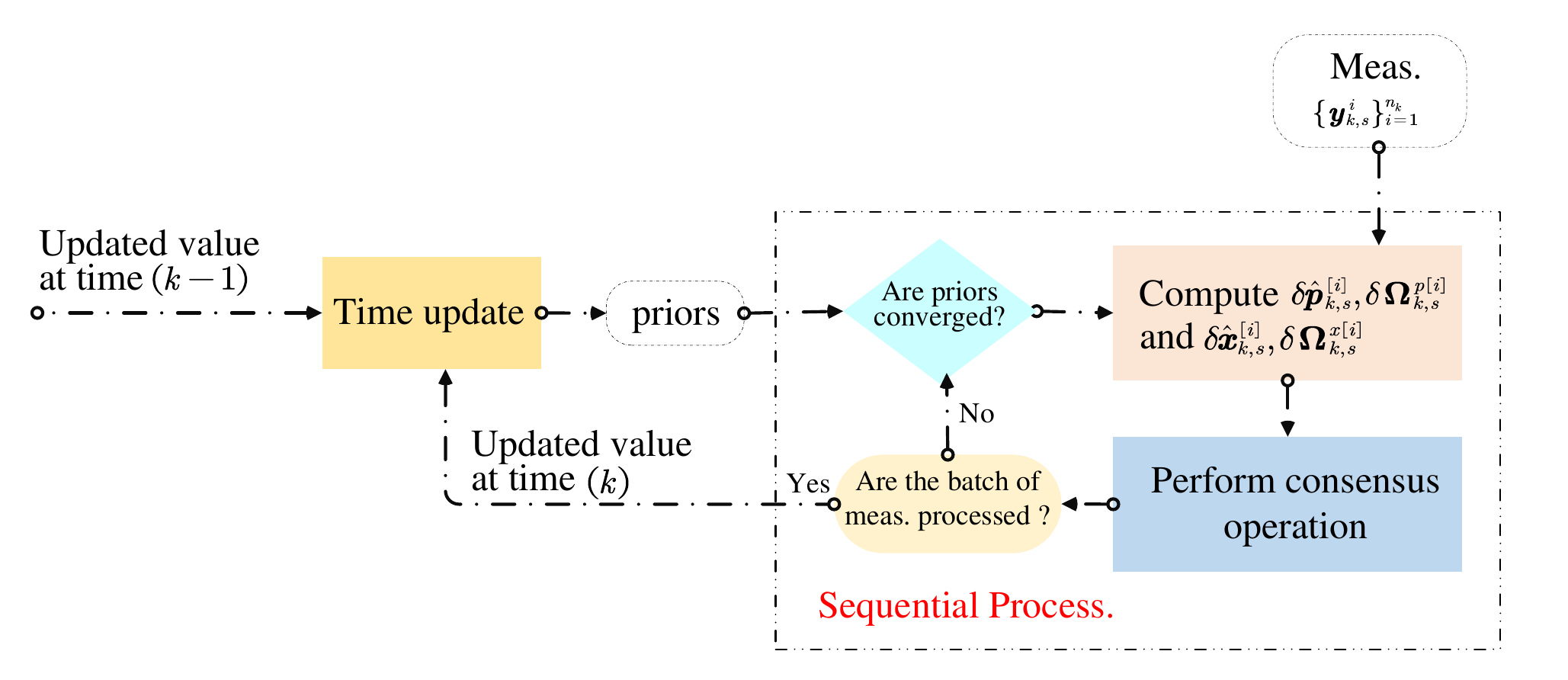}.
	{\color {red} {	\caption{\small{Block diagram of the proposed DWLSF.}}}}
	\label{loop}
\end{figure*}

The DWLSF involves the temporal evolution, sequential processing, and consensus iteration blocks (for detail, see Fig. 4). The detailed DWLSF is collected in Algorithm \ref{algorithm2}. 
\begin{algorithm}[h]
	\caption{DWLSF Filter}\label{algorithm2}	
	\SetAlgoLined
	\textbf{Initialization:} $\hat{\bm x}_{1,s}^{[0]}$, $\hat{\bm p}_{1,s}^{[0]}$, $\mathbf{\Omega}_{1,s}^{x[0]}$, and $\mathbf{\Omega}_{1,s}^{p[0]}$ \;
	\For{$k \gets 1,2,\cdots $ \tcp*[h]{scan time} }{
		\textbf{Data:} $\{\bm y_{k,s}^{i} \}^{n_{k}}_{i=1} (s \in \mathcal N)$ \;
		Initialization: $\hat{\bm x}_{k,s}^{[0]}$, $\hat{\bm p}_{k,s}^{[0]}$, $\mathbf{\Omega}_{k,s}^{x[0]}$, and $\mathbf{\Omega}_{k,s}^{p[0]}$ \;
		
		\For{$i=1,2,\cdots,n_k$ \tcp*[h]{sequential} }{\textbf{Compute consensus quantities} \; 
			\eIf{Converged priors}{using \eqref{eq43} and \eqref{eq45a}, compute $\delta \hat{\bm x}^{[i]}_{k,s}, \delta \mathbf{\Omega}^{x[i]}_{k,s}, \delta \hat{\bm p}^{[i]}_{k,s}, \delta \mathbf{\Omega}^{p[i]}_{k,s}$ via \eqref{eq38} and \eqref{eq39}}{using \eqref{eq44} and \eqref{eq45b}, compute $\delta \hat{\bm x}^{[i]}_{k,s}, \delta \mathbf{\Omega}^{x[i]}_{k,s}, \delta \hat{\bm p}^{[i]}_{k,s}, \delta \mathbf{\Omega}^{p[i]}_{k,s}$ via \eqref{eq38} and \eqref{eq39}}
		\textbf{Consensus operation} \;
		set $\delta \hat{\bm x}^{[i]}_{k,s}(0) \gets\delta \hat{\bm x}^{[i]}_{k,s}$ , $\delta \mathbf{\Omega}^{x[i]}_{k,s}(0) \gets \delta \mathbf{\Omega}^{x[i]}_{k,s}$,  $\delta \hat{\bm p}^{[i]}_{k,s}(0)\gets\delta \hat{\bm p}^{[i]}_{k,s}$, $\delta \mathbf{\Omega}^{p[i]}_{k,s}(0) \gets\delta \mathbf{\Omega}^{p[i]}_{k,s}$ 	\;
		\For{$l = 0,\cdots,L-1$}{perform AC on $\delta \hat{\bm x}^{[i]}_{k,s}, \delta \mathbf{\Omega}^{x[i]}_{k,s}, \delta \hat{\bm p}^{[i]}_{k,s}, \delta \mathbf{\Omega}^{p[i]}_{k,s}$}
		compute $\hat{\bm x}_{k,s}^{[i]},\mathbf{\Omega}_{k,s}^{x[i]},\hat{\bm p}_{k,s}^{[i]},\mathbf{\Omega}_{k,s}^{p[i]}$ via \eqref{eq40} and \eqref{eq41}
	}
		\textbf{Output1:} $\hat{\bm{x}}_{k,s} \gets \hat{\bm x}_{k,s}^{[n_k]}, \mathbf C_{k,s}^{x} \gets \left( \mathbf{\Omega}_{k,s}^{x[n_k]}\right)^{-1}$ \;
		\textbf{Output2:} $\hat{\bm{p}}_{k,s} \gets \hat{\bm p}_{k,s}^{[n_k]},\mathbf C_{k,s}^{p} \gets \left( \mathbf{\Omega}_{k,s}^{p[n_k]}\right)^{-1}$ \;
		compute $\hat{\bm x}_{k+1,s}^{[0]},\mathbf{\Omega}_{k+1,s}^{x[0]},\hat{\bm p}_{k+1,s}^{[0]},\mathbf{\Omega}_{k+1,s}^{p[0]}$ via \eqref{eq24} and \eqref{eq25}	
	}
\end{algorithm}

 \begin{remark}
 	\begin{itemize}
 		\item The number of iterations $L$ linearly increases both computation and communication burdens. Hence, its value is usually set to be a suitable value to balance the performance and cost, which renders the estimates on each node being asymptotically optimal.
 		\item At the initial scan time $k=1$, the uncorrelated priors (i.e., \eqref{eq44} and \eqref{eq45b}) are used in the sequential processing process. Then for $k \ge 2$, \eqref{eq43} and \eqref{eq45a} should be used as the estimates across the nodes become correlated for the following time steps \cite{b27}.
 		\item If the priors converge, the information becomes redundant on each node, and thus dividing the information matrices by ${\vert \mathcal N \vert}$ is necessary to match the centralized estimates. 
 	\end{itemize}
 \end{remark}

\section{Numerical Examples}
\label{sec:simulation}
To evaluate the performance of CVM and MEM proposed in \cite{b11}, we first integrate the two models into CWLSF to track a rectangular object over a network with single node. Then, under different prior conditions, we compare DWLSF with the two previous approaches in \cite{Li21} and \cite{bb} (abbreviated as CI filter and DEOT filter, respectively) in terms of the estimate consensus, computation time and tracking error over a network including the ``naive'' nodes. All numerical simulations are operated in MATLAB--2019b running on a PC with processor \texttt {Intel(R) Core(TM) i7-10510U CPU @ 1.8GHz 2.3GHz and with 20GB RAM.} 
\subsection{Performance evaluation on CVM and MEM model}
To examine the effectiveness of CVM and MEM on different tracking applications, we conduct two experiments: 1) the orientation is aligned to the direction of velocity; 2) the orientation does not move exactly along the direction of velocity. We assess the position and extent errors simultaneously by the Optimal Sub-Pattern Assignment (OSPA) distance \cite{b10}.
\subsubsection{Object moves along the orientation}
\begin{figure}[htbp]
	\centering
	\includegraphics[scale=0.73]{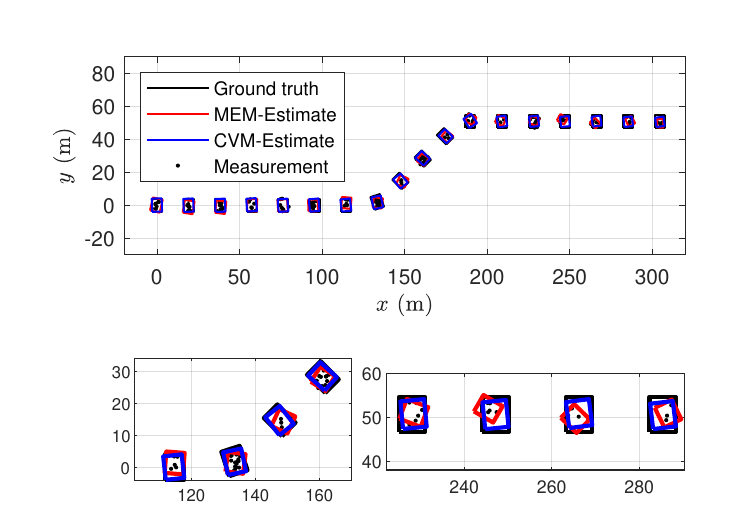}
	\caption{\small{True trajectory and overall tracking results (per 3 scan times)}} 
	\label{fig5}
\end{figure}

\begin{figure} [htbp]
	\centering
	\includegraphics[scale=0.6]{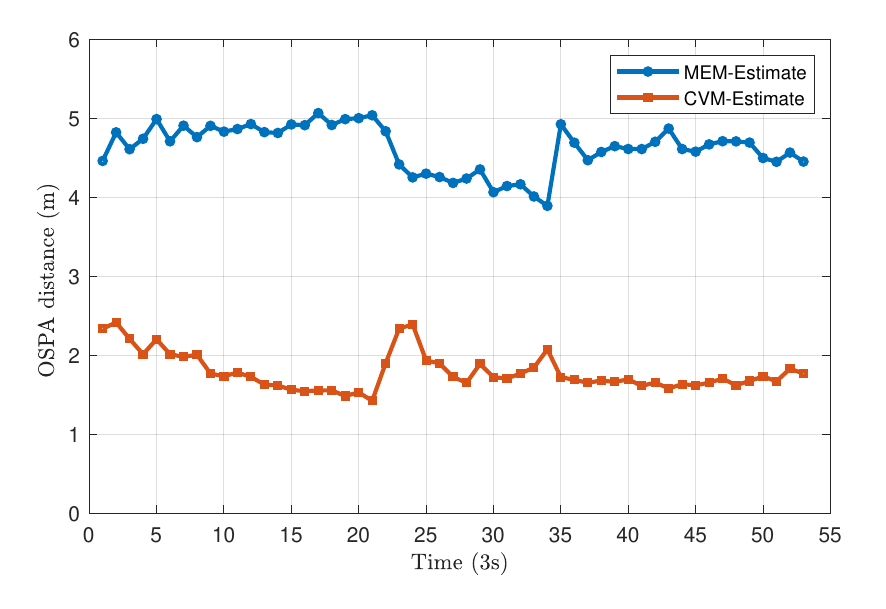}
	\caption{\small{OSPA distance}}
	\label{fig6}
\end{figure}

In this scenario (S1), the EO is a rectangular object with lengths $3$ and $4$ meters. The object moves with a nearly constant velocity $v=1.5 \mathrm{m/s}$ following the trajectory shown in Fig. \ref{fig5}. The initial position of the object is at the origin of coordinates, and its orientation is consistent with the direction of velocity. Table \ref {tab:1} collects the parameters used in CWLSFs. 
 \begin{table}[htpb]
	\renewcommand\arraystretch{1.4}
	\centering
	\setlength\tabcolsep{0.45pt}
	\caption{Parameter Setting in S1}\label{tab:1}
	\begin{tabular}{c|cc}
		\Xhline{1pt}
		Categories & Para. & Specification \\ \hline
		\multirow{7}*{Common para.} & Scan time & $\mathrm{T}=3$ s \\
		& Mea. Cov. & $\mathbf C^v = \mathrm{diag}(\frac{1}{3},\frac{1}{3})$ \\
		& Multip. noise Cov. & $\mathbf C^h =  \frac{1}{3} \mathbf I_2$ \\
		& Kine. transition matrix & $\mathbf F_k^s = \begin{bmatrix}
			1  & 0 & \mathrm{T} & 0 \\ 0 & 1 & 0 & \mathrm{T} \\ 0 & 0 & 1 & 0 \\ 0 & 0 & 0 & 1 
		\end{bmatrix}$ \\
		& Process Cov. in Kine. & $\mathbf C^x_w = \mathrm{diag}(50,50,1,1)$ \\
		& Cov. in Extent & $\mathbf {C}_{1}^{p[0]} = \mathrm{diag} (0.36, \frac{1}{500}, \frac{1}{50})$ \\	
		& Cov. in Kine. & $\mathbf {C}_{1}^{x[0]} = \mathrm{diag} (2,2, \frac{1}{5}, \frac{1}{5})$ \\			
		& Extent transition matrix & $\mathbf F_k^p = \mathbf I_3$ \\	
		& No. of Meas. & $\lambda =7 $ \\ \hline
		\multirow{1}*{MEM} 	
		& Process Cov. in Extent & $\mathbf C^p_w = \mathrm{diag}(0.05, \frac{1}{500}, \frac{1}{50})$ \\	 \hline	
		\multirow{1}*{CVM} 		
		& Process Cov. in Extent & $\mathbf C^p_w = \mathrm{diag}(0.3, \frac{1}{500}, \frac{1}{220})$ \\	
		\Xhline{1pt}
	\end{tabular}
\end{table}
Fig. \ref{fig5} gives the true trajectory and overall tracking results over $M=50$ Monte Carlo runs. As shown in Fig. \ref{fig5}, CWLSF-CVM has better precision in comparison with CWLSF-MEM, especially during the turning phase and final moving phase. This because CVM indeed describes the tight relation between the velocity and orientation, so that when the object's motion pattern changes, such as turning maneuvering, CWLSF-CVM quickly captures the change to modify its filtering gain. Fig. \ref{fig6} gives the OSPA distance, and the result provides a direct conclusion on the advantage of CWLSF-CVM. 

\subsubsection{Object moves with a drift}
\begin{figure}[htbp]
	\centering
	\includegraphics[scale=0.73]{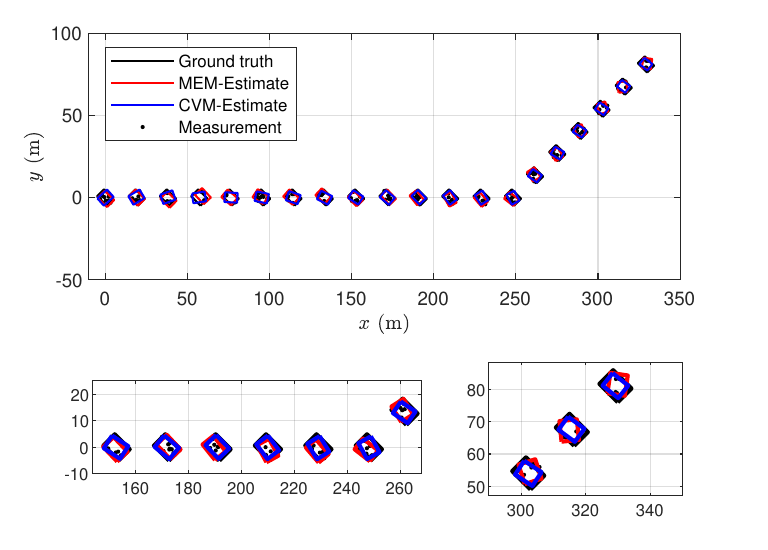}
	\caption{\small{True trajectory and overall tracking results (per 3 scan times)}} 
	\label{fig7}
\end{figure}

In this scenario (S2), the EO is a rectangular object with lengths $3$ and $4$ meters. The object moves with a nearly constant velocity $v=1.5 \mathrm{m/s}$ following the trajectory shown in Fig. \ref{fig7}. The initial position of the object is at the origin of coordinates, and its orientation is a constant value $\frac{\pi}{4}$. Table \ref {tab:2} collects the parameters used in S2, and the other parameters are given in Table \ref {tab:1}. 
\begin{table}[htpb]
		\renewcommand\arraystretch{1.4}
	\centering
	\setlength\tabcolsep{0.45pt}
	\caption{Parameter Setting in S2}\label{tab:2}
	\begin{tabular}{c|cc}
		\Xhline{1pt}
		Categories & Para. & Specification \\ \hline
		\multirow{1}*{Common para.} 
		& Cov. in Extent & $\mathbf {C}_{1}^{p[0]} = \mathrm{diag} (0.0.1, \frac{1}{500}, \frac{1}{100})$ \\	 \hline
		\multirow{1}*{MEM} 	
		& Process Cov. in Extent & $\mathbf C^p_w = \mathrm{diag}(0.015, \frac{1}{400}, \frac{1}{300})$ \\	 \hline	
		\multirow{1}*{CVM} 		
		& Process Cov. in Extent & $\mathbf C^p_w = \mathrm{diag}(0.5, \frac{1}{400}, \frac{1}{300})$ \\	
		\Xhline{1pt}
	\end{tabular}
\end{table}

\begin{figure} [htbp]
	\centering
	\includegraphics[scale=0.6]{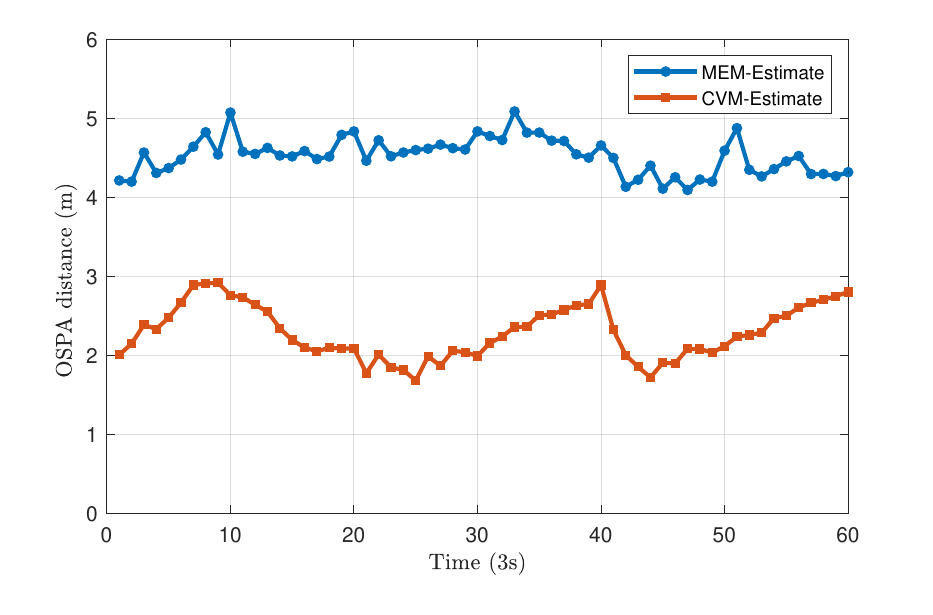}
	\caption{\small{OSPA distance}}
	\label{fig8}
\end{figure}

Fig. \ref{fig7} gives the true trajectory and overall tracking results over $M=50$ Monte Carlo runs. As expected, CWLSF-CVM outperforms CWLSF-MEM as a whole. The reason is that CVM allows objects to do a drift motion, i.e., a mismatch between the orientation and direction of their velocity, while MEM assumes that objects move along their orientation. Fig. \ref{fig8} shows that CWLSF-CVM has lower OSPA distance than that of CWLSF-MEM, which validates the superiority on CVM.

\subsection{Performance evaluation in a distributed scenario}
\begin{figure}[htpb]
	\centering
	\includegraphics[scale=0.7]{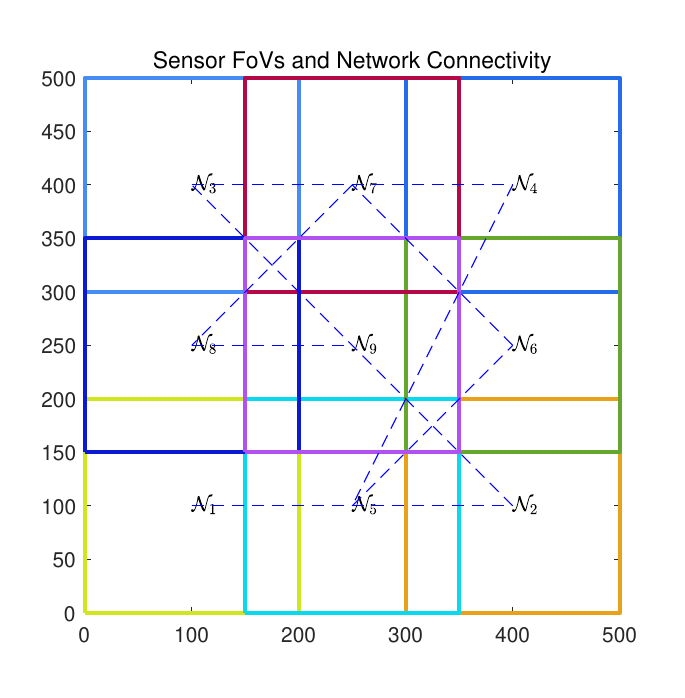}
	\caption{\small{FoVs and network connectivity. The sensor FoVs are shown in squares. There are $9$ sensors in this example. The blue dotted lines represent the network connectivity.}}
	\label{fig9}
\end{figure}

In this scenario (S3), the network deploys $9$ nodes to monitor a $[0,500] \mathrm{m} \times [0,500] \mathrm{m}$ space. The sensing range is $200$m, and sensing azimuth spans from $0^{\circ}$ to $360^{\circ}$ for all sensors. Fig. \ref{fig9} gives an example for such a network. Here, the considered EO is an elliptical object with lengths of the semi-axes $35$m and $30$m. The initial position of the object is at $[25,300] $, and then moves with a nearly constant velocity $v=\frac{100}{36} \mathrm{m/s}$ following a similar trajectory as in \cite{b28}. A sensor has measurements of the object only if the ground truth position of the object is within the sensor’s FoV. 

Since DWLSF is guaranteed to converge to the centralized estimates only if the initial priors are equal, we testify the robustness of DWLSF in three cases: the equal priors, uncorrelated and unequal priors, and correlated and unequal priors. The comparison results are the Gaussian Wasserstein distance (GWD) \cite{b10,b11} for assessing both the position and extent errors, and a metric, Averaged consensus estimate error (ACEE), for testifying the estimate difference between different nodes. Moreover, considering that the number of iterations has a critical impact on filters' performance, we check how many iterations $L$ could achieve a stable behavior.

\subsubsection{Equal priors}
If the priors are equal for all nodes, which means the priors have converged at the initial scan time $k=1$. Thus, the prior covariances w.r.t kinematics are set to $\mathbf {C}_{1,s}^{x[0]} = \mathrm{diag} (50,50,10,10)$, and the prior covariances w.r.t extent are set to $\mathbf {C}_{1,s}^{p[0]} = \mathrm{diag} (0.01,0.1,0.1)$ for all nodes. The other parameters are given in Table \ref{tab:3}.

\begin{table}[htpb]
		\renewcommand\arraystretch{1.4}
	\centering
	\setlength\tabcolsep{0.6pt}
	\caption{Parameter Setting in S3}\label{tab:3}
	\begin{tabular}{c|cc}
		\Xhline{1pt}
		Categories & Para. & Specification \\ \hline
		\multirow{11}*{Common para.} & Scan time & $\mathrm{T}=5$ s \\
		& Mea. Cov. & $\mathbf C^v = \mathrm{diag}(40,20)$ \\
		& Multip. noise Cov. & $\mathbf C^h =  \frac{1}{4} \mathbf I_2$ \\
		& Kine. transition matrix & as shown in Table \ref{tab:1} \\
		& Process Cov. in Kine. & $\mathbf C^x_w = \mathrm{diag}(10^{2},10^{2},1,1)$ \\
		& Prior cov. in CWLSF & $\mathbf {C}_{1}^{x[0]} = \mathrm{diag} (2,2, \frac{1}{2},\frac{1}{2})$ \\
		& Prior cov. in CWLSF & $\mathbf {C}_{1}^{p[0]} = \mathrm{diag} (2\times 10^{-3},10^{-3} , 10^{-4})$  \\	
		& Consensus para. & $\xi = 0.65/\triangle\max$ \\
		& Maximum degree & $\triangle\max =2$ \\	
		& Extent transition matrix & $\mathbf F_k^p = \mathbf I_3$ \\	
		& No. of Meas. & $\lambda =10 $ \\ \hline
		\multirow{1}*{Others filter} 	
		& Process Cov. in Extent & $\mathbf C^p_w = \mathrm{diag}(0.05 , 10^{-3} , 10^{-4})$ \\	 \hline	
		\multirow{1}*{CVM} 	
		& Process Cov. in Extent & $\mathbf C^p_w = \mathrm{diag} (2\times 10^{-3},10^{-3} , 10^{-4})$  \\		
		\Xhline{1pt}
	\end{tabular}
\end{table}
\begin{figure}[htbp]
	\begin{minipage}[t]{0.5\textwidth}
		\centering\includegraphics[scale=0.6]{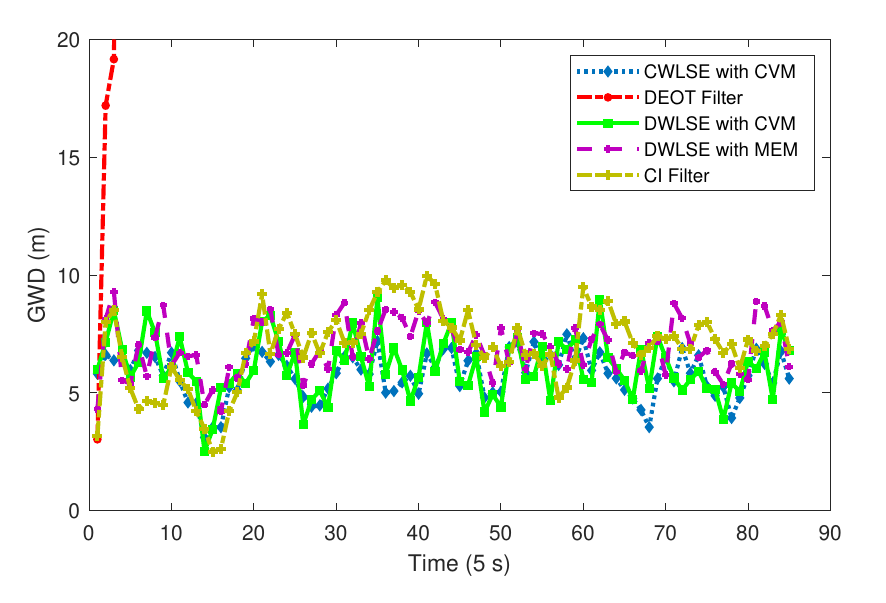}
		\caption{\small{GWD distance with $L=10$}}
		\label{fig10}
	\end{minipage}
	\begin{minipage}[t]{0.5\textwidth}
		\centering\includegraphics[scale=0.6]{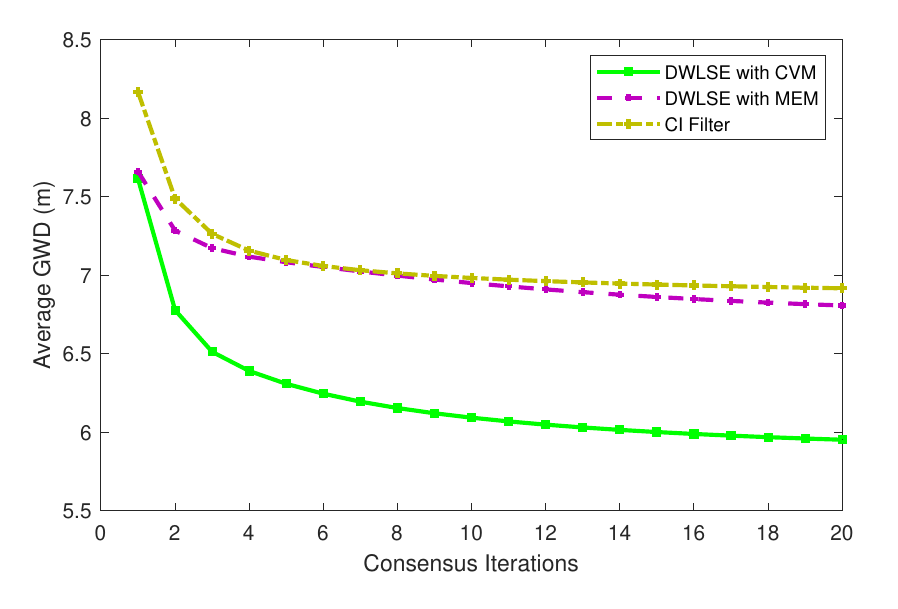}
		\caption{\small{Average GWD distance with different iterations}}
		\label{fig11}
	\end{minipage}
\end{figure}

\begin{figure}[htbp]
	\begin{minipage}[t]{0.5\textwidth}
		\centering\includegraphics[scale=0.6]{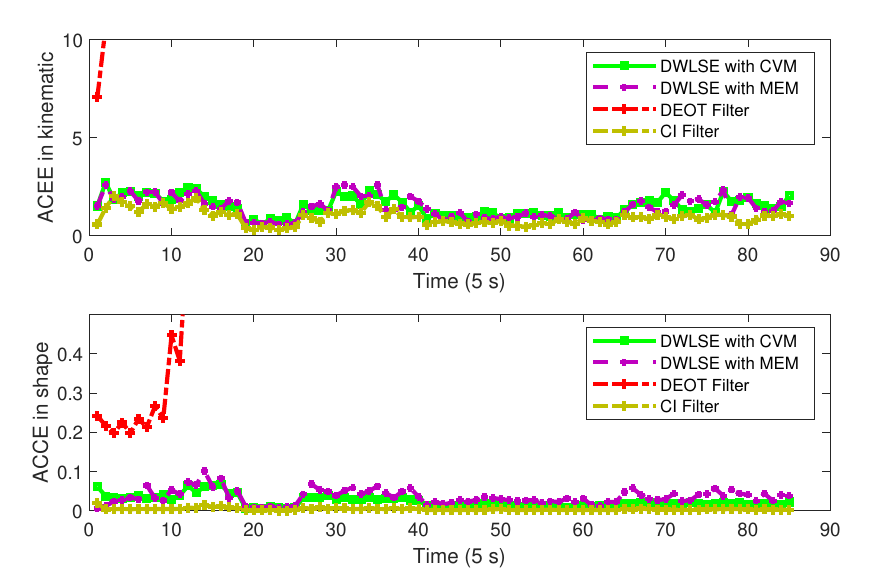}
		\caption{\small{ACEEs in kinematics and extent ($L=10$)}}
		\label{fig12}
	\end{minipage}
	\begin{minipage}[t]{0.5\textwidth}
		\centering\includegraphics[scale=0.6]{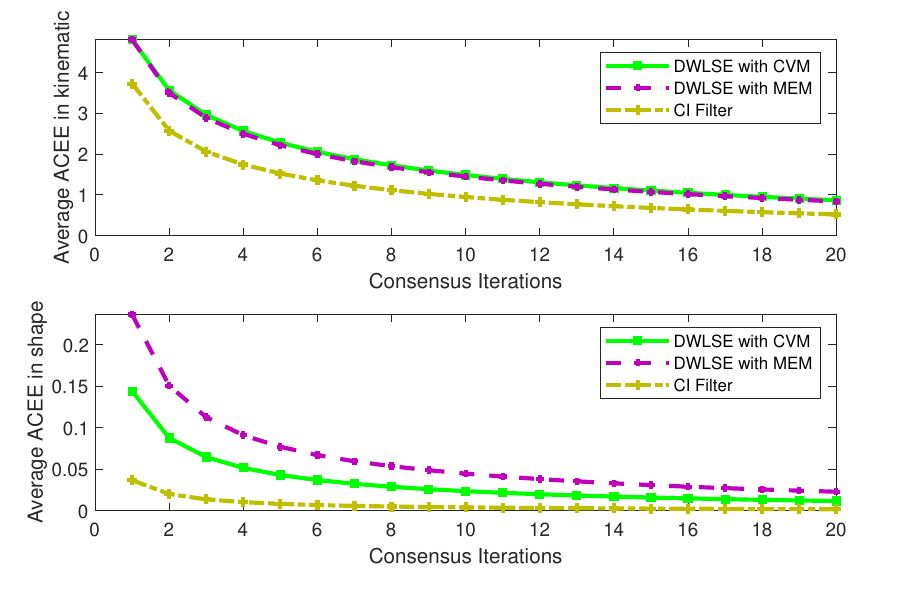}
		\caption{\small{Average ACEEs with different iterations}}
		\label{fig13}
	\end{minipage}
\end{figure}

 \begin{figure}[t]
	\centering
	\includegraphics[scale=0.6]{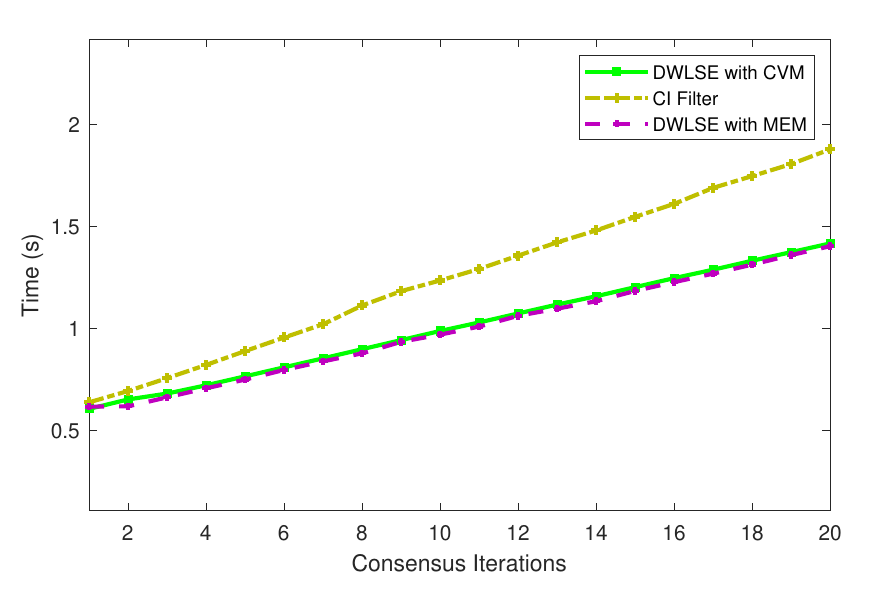}
	\caption{\small{Computational time with different iterations.}}
	\label{fig14}
\end{figure}
Fig. \ref{fig10} shows the GWD distance of four examined filters. The DWLSF-CVM performs better than the other distributed filters, as CVM delivers the merit that describes the correlation between the velocity and orientation to DWLSF-CVM. DEOT filter is divergent in S3, because it assumes that each node in network detects the object during the whole tracking process. However, the assumption is no longer valid in a realistic network as in S3. The reason why CI filter has a lager GWD distance than DWLSFs is that the cross-covariances across different nodes are not incorporated in its estimation framework.

As expected, the average GWD distance decreases with the increased number of iterations, and this phenomenon is more apparent in DWLSF-CVM (see Fig. \ref{fig11}). Combined the results in Figs. \ref{fig10}-\ref{fig11}, we can anticipate that DWLSF-CVM will approach to CWLSF-CVM almost tightly when $L>10$, which verifies the validity of the proposed consensus scheme.

Fig. \ref{fig12} shows the difference between nodes in the kinematics and extent estimate. Both DWLSFs give a satisfied consensus result in the kinematics, but DWLSF-CVM has a smaller difference in the extent. Although CI filter has a lower ACEE, one cannot declare that it outperforms DWLSEs. Because when the GWD distance is large, even if the ACEE is small, it does not make any sense. Again, DEOT filter fails to yield a satisfied consensus result.

Fig. \ref{fig13} shows the average ACEE under different iterations. Compared with DWLSF-MEM, DWLSF-CVM and CI filter require fewer iterations to generate a stable consensus result. 

Fig. \ref{fig14} compares the average computational costs per tracking process on different filters under different iterations. For a given $l$, DWLSFs almost consume $70\%$ computational resource of CI filter. This merit makes DWLSFs more appealing on a computer with limited computing capability.

\subsubsection{Uncorrelated and unequal priors}
\label{sub:VI-B-2}
In this case, the prior covariances w.r.t kinematics are set to $\mathbf {C}_{1,s}^{x[0]} = \mathrm{diag} (100, 100, 10, 10).*\mathrm{diag}(\mathrm{rand}(1,4))$, and the prior covariances w.r.t extent are set to $\mathbf {C}_{1,s}^{p[0]} = \mathrm{diag} (1,7,7).*\mathrm{diag}(\mathrm{rand}(1,3))$ for all nodes. The other parameters are given in Table \ref{tab:3}. 

\begin{figure}[htbp]
	\begin{minipage}[t]{0.5\textwidth}
		\centering\includegraphics[scale=0.6]{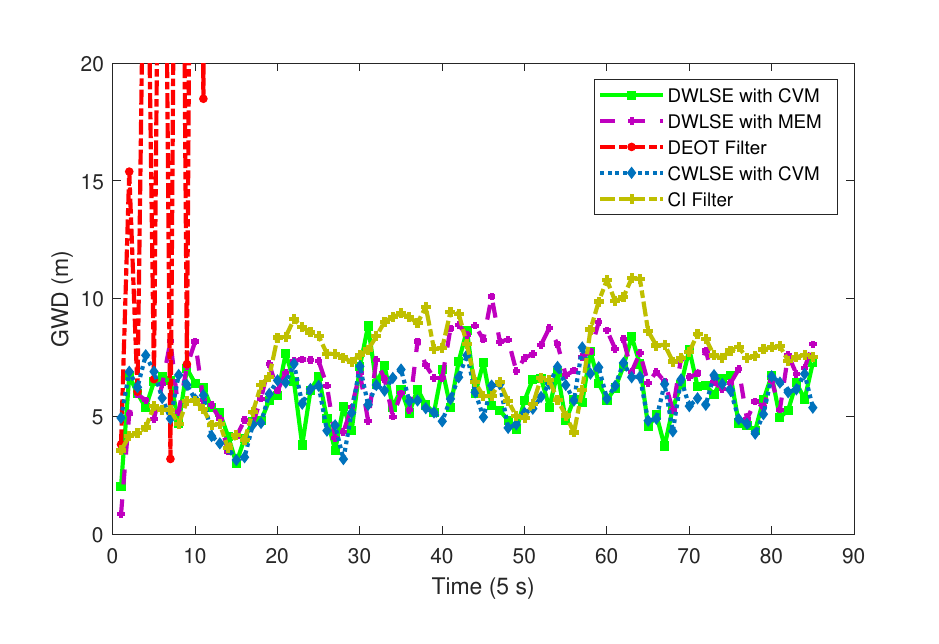}
		\caption{\small{GWD distance with $L=10$}}
		\label{fig15}
	\end{minipage}
	\begin{minipage}[t]{0.5\textwidth}
		\centering\includegraphics[scale=0.6]{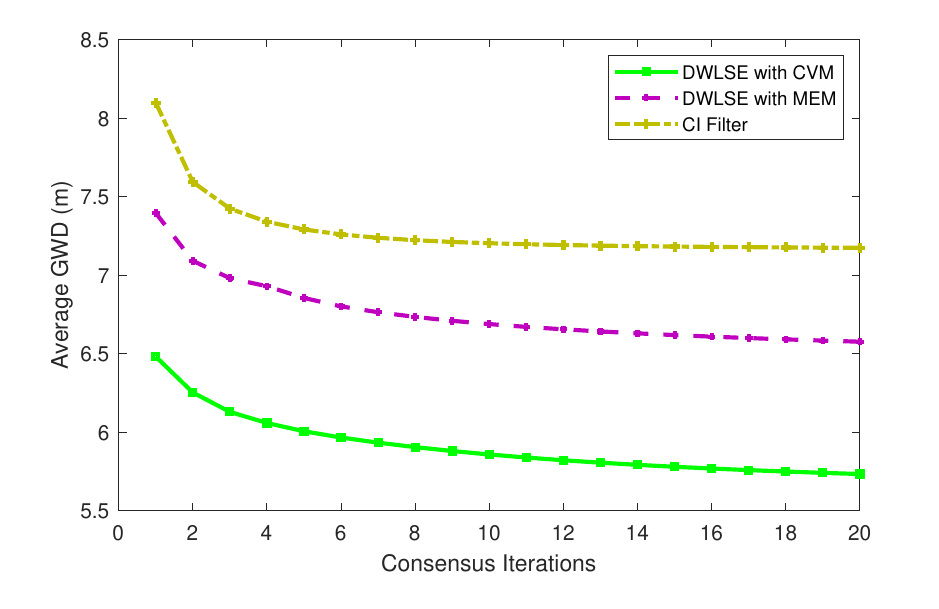}
		\caption{\small{Average GWD distance with different iterations}}
		\label{fig16}
	\end{minipage}
\end{figure}

As shown in Figs. \ref{fig15}-\ref{fig16}, even in the case that the priors are unequal and uncorrelated, DWLSF-CVM still follows CWLSF-CVM closely with increased iterations. This because DWLSF is a consensus-based filter and irrespective of the initial condition, after several time steps or iterations, the priors have met the converged condition. Also, DWLSF-CVM has minimum GWD distance in comparison with the other distributed filters under the case. 

\begin{figure}[htbp]
	\begin{minipage}[t]{0.5\textwidth}
		\centering\includegraphics[scale=0.6]{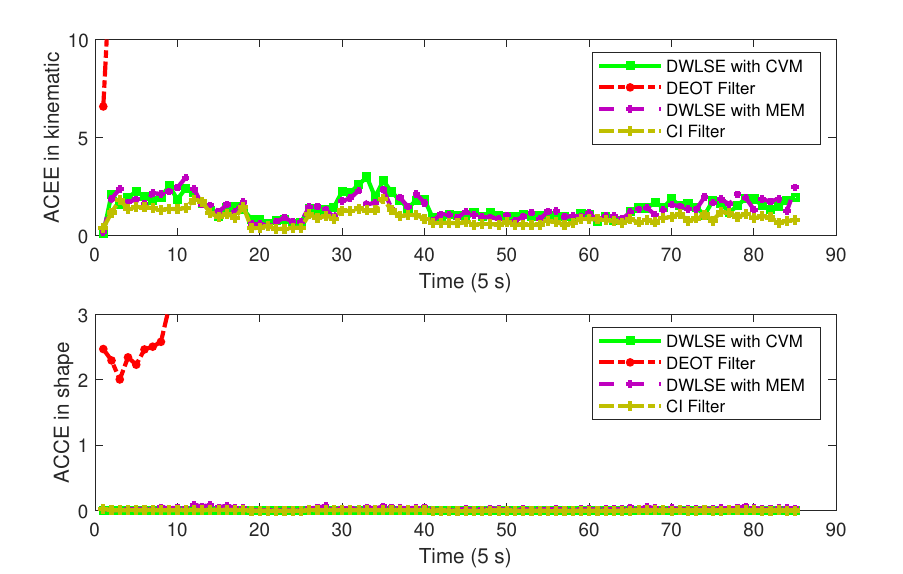}
		\caption{\small{ACEEs in kinematics and extent ($L=10$)}}
		\label{fig17}
	\end{minipage}
	\begin{minipage}[t]{0.5\textwidth}
		\centering\includegraphics[scale=0.6]{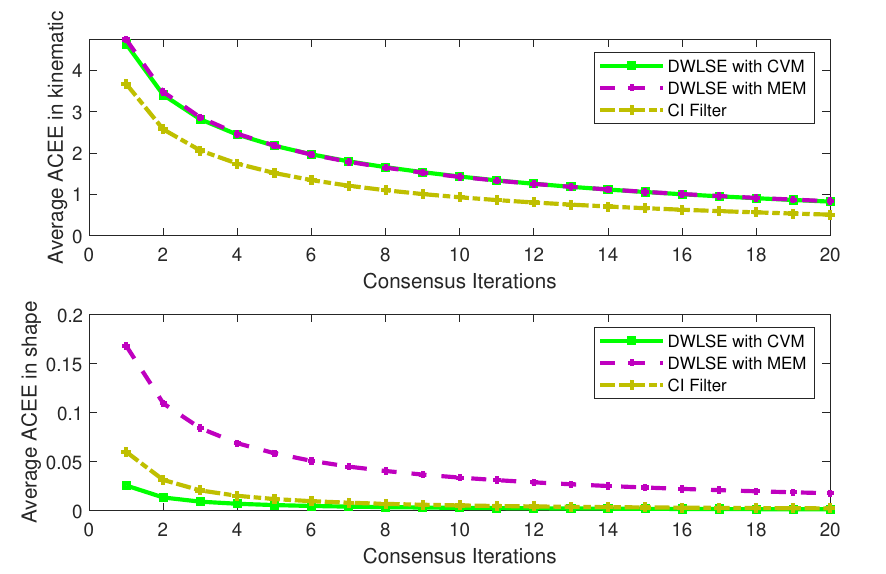}
		\caption{\small{Average ACEEs with different iterations}}
		\label{fig18}
	\end{minipage}
\end{figure}

For the estimate difference between nodes, both DWLSFs give a satisfying consensus result as shown in Figs. \ref{fig17}-\ref{fig18}. These results prove the robustness of DWLSF in terms of the consensus. The ACEEs on CI filter are similar to DWLSFs' result, while the ACEEs on DEOT are beyond a reasonable range. 

\begin{figure}[htbp]
	\centering
	\includegraphics[scale=0.6]{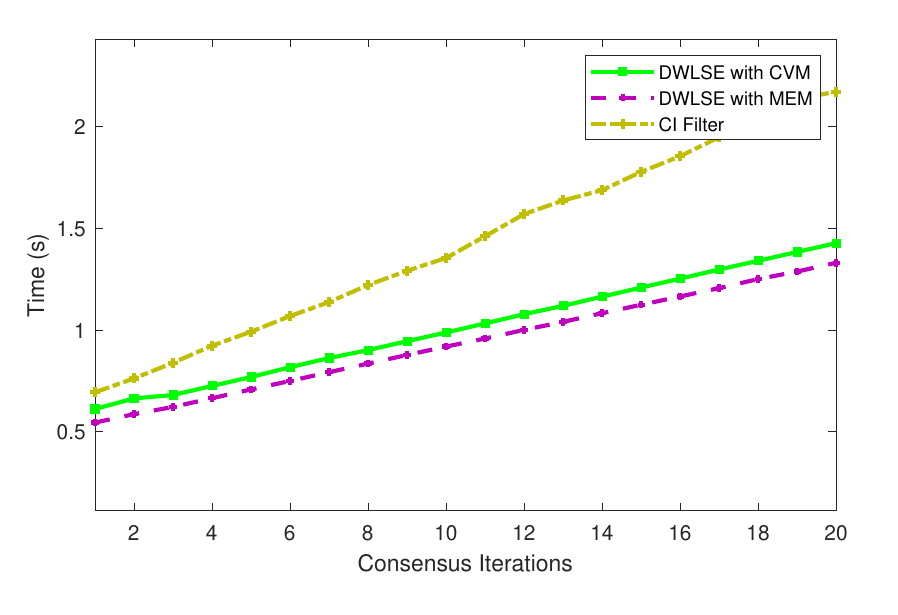}
\caption{\small{Computational time with different iterations.}}
	\label{fig19}
\end{figure}

The average computational costs per tracking process under different iterations are given in  Fig. \ref{fig19}. Combined with the results in Figs. \ref{fig15}-\ref{fig16}, we see that DWLSFs require less computational resources to yield a pretty performance in comparison with CI filter.

\subsubsection{Correlated and unequal priors}
In this case, the prior covariances w.r.t kinematics are set to $\mathbf {C}_{1,s}^{x[0]} = \mathrm{diag} (100, 100, 10, 10).*\mathrm{diag}(\mathrm{rand}(1,4))$, and the prior covariances w.r.t extent are set to $\mathbf {C}_{1,s}^{p[0]} = \mathrm{diag} (1,7,7).*\mathrm{diag}(\mathrm{rand}(1,3))$, with $\rho=0.5$ correlation coefficient between the priors across nodes. The other parameters are given in Table \ref{tab:3}. 
\begin{figure}[htbp]
	\begin{minipage}[t]{0.5\textwidth}
		\centering\includegraphics[scale=0.6]{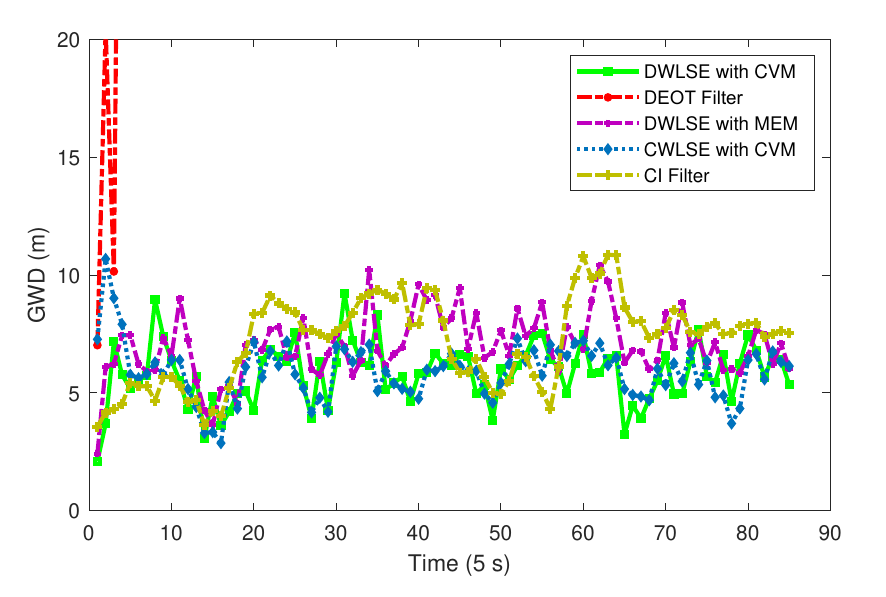}
		\caption{\small{GWD distance with $L=10$}}
		\label{fig20}
	\end{minipage}
	\begin{minipage}[t]{0.5\textwidth}
		\centering\includegraphics[scale=0.6]{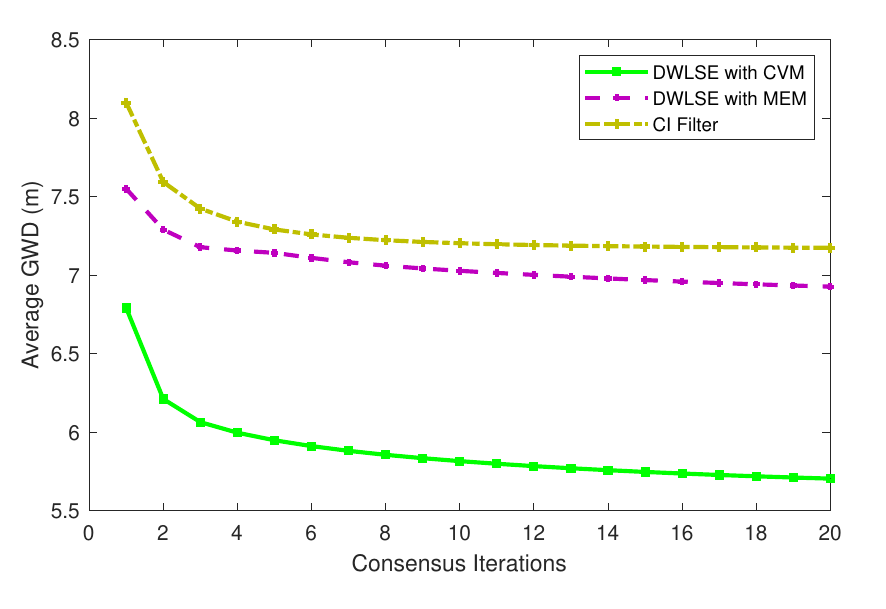}
		\caption{\small{Average GWD distance with different iteration}}
		\label{fig21}
	\end{minipage}
\end{figure}

\begin{figure}[h]
		\centering\includegraphics[scale=0.6]{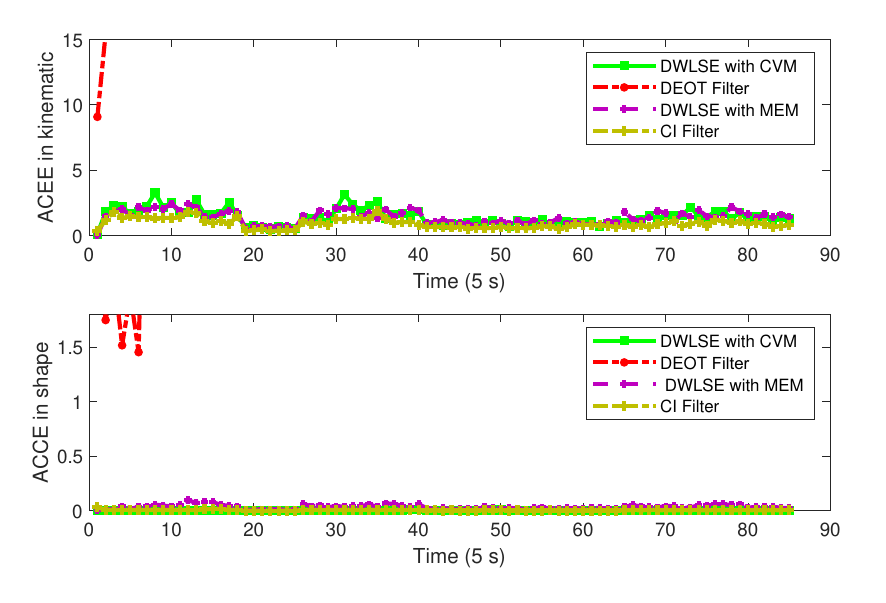}
		\caption{\small{ACEEs in kinematics and extent ($L=10$)}}
		\label{fig22}
\end{figure}

\begin{figure}[h]
		\centering\includegraphics[scale=0.6]{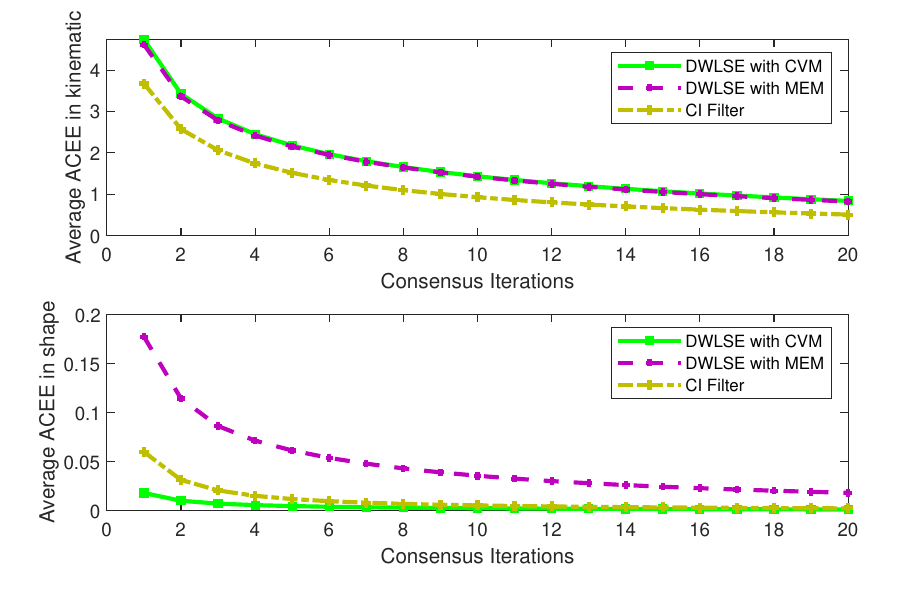}
		\caption{\small{Average ACEEs with different iterations}}
		\label{fig23}
\end{figure}

 \begin{figure}[t]
	\centering
	\includegraphics[scale=0.6]{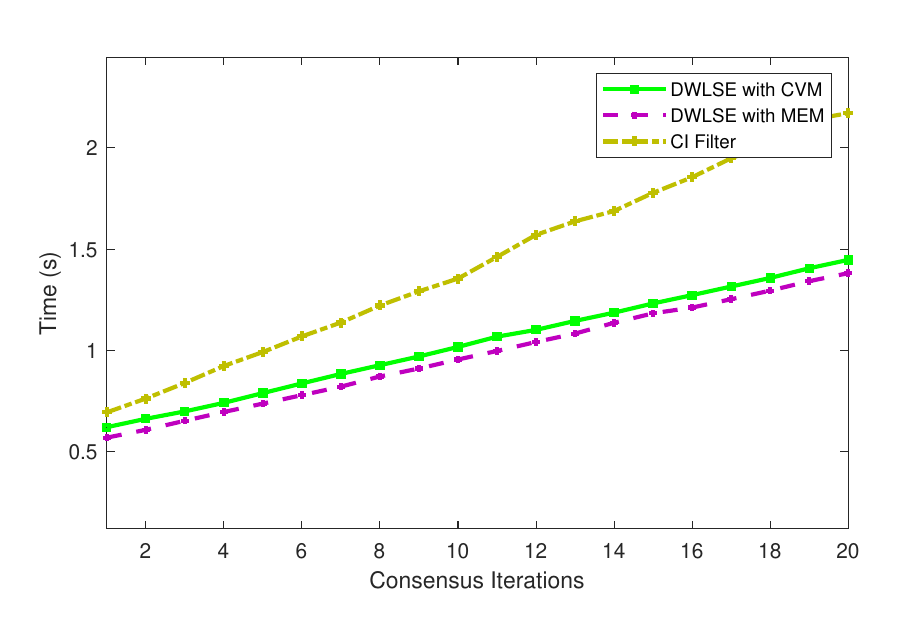}
	\caption{\small{Computational time with different iterations.}}
	\label{fig24}
\end{figure}

From the results in Figs. \ref{fig20}-\ref{fig21}, DWLSE-CVM also approaches to the result of CWLSE-CVM when the priors are unequal and correlated. As for the ACEE, average ACEE and computation time (see Figs. \ref{fig22}-\ref{fig24}), the same conclusion is drawn as given in section \ref{sub:VI-B-2}.
\section{Conclusion}
\label{sec:conclusion}
This study first proposes a coupled velocity state-space model. This model is capable of describing the correlation between the object's orientation and velocity, which allows the model to further detect changes in the EO's motion mode and improve the performance of the corresponding filter. Then, without losing any first and second moment information, the proposed model is separated into two pseudo-linearized models with only additive noise. Finally, under a network with ``naive" nodes, we use the two models to design a distributed WLS filter that takes the cross-covariances between nodes into account. The distributed filter converges to the corresponding centralized form in three different situations, which indicates its robustness when the convergence condition is not met. Potential future works could ponder how to achieve the distributed filter in a heterogeneous sensor network \cite{b30,b31} or under multiple constraints \cite{b32,b33,xu2022}.

%
%
%
\section*{Acknowledgment}
This work is supported by the National Natural Science
Foundation of China (Grant no. 61873205 and 61771399).
%
\appendices
\section{Proof of proposition \ref{the1}} \label{proof1}
\begin{IEEEproof}
	For brevity, we first derive the partial derivatives of $\frac{v_k^{\mathsf{x}}}{\Vert \bm {\vartheta}_k \Vert}$ and $\frac{v_k^{\mathsf{y}}}{\Vert \bm {\vartheta}_k \Vert}$ with respect to $v_k^{\mathsf{x}}$ and $ v_k^{\mathsf{y}}$, respectively, at the $[i-1]$-th velocity estimate $ {{\bm \vartheta}_k^{[i-1]}}$ as 
	
	\begin{equation}\label{eq46}
		\left \{ \begin{lgathered} 
			d1 = \left. \frac{\partial v_k^{\mathsf{x}} / \Vert \bm {\vartheta}_k \Vert}{\partial v_k^{\mathsf{x}} } \right|_{\hat{\bm \vartheta}_k^{[i-1]}} = \left. \frac{1}{\Vert \bm {\vartheta}_k \Vert} - \frac{(v_k^{\mathsf{x}})^2}{(\Vert \bm {\vartheta}_k \Vert)^3} \right|_{\hat{\bm \vartheta}_k^{[i-1]}},
			\\
			d2 = \left. \frac{\partial v_k^{\mathsf{y}}/ \Vert \bm {\vartheta}_k \Vert}{\partial v_k^{\mathsf{y}} } \right|_{\hat{\bm \vartheta}_k^{[i-1]}} = \left. \frac{1}{\Vert \bm {\vartheta}_k \Vert} - \frac{(v_k^{\mathsf{y}})^2}{(\Vert \bm {\vartheta}_k \Vert)^3} \right|_{\hat{\bm \vartheta}_k^{[i-1]}},	\\
			d3 = \left. \frac{\partial v_k^{\mathsf{y}}/ \Vert \bm {\vartheta}_k \Vert}{\partial v_k^{\mathsf{x}} } = \frac{\partial v_k^{\mathsf{x}}/ \Vert \bm {\vartheta}_k \Vert}{\partial v_k^{\mathsf{y}} } \right|_{\hat{\bm \vartheta}_k^{[i-1]}} = \left. - \frac{v_k^{\mathsf{x}} v_k^{\mathsf{y}}}{(\Vert \bm {\vartheta}_k \Vert)^3} \right|_{\hat{\bm \vartheta}_k^{[i-1]}}.
		\end{lgathered} \right.	
	\end{equation}
	
	Considering that the true extent $\bm p_k$ and velocity $\bm {\vartheta}_k$ are unknown in the coefficient matrix $\mathbf S_k$, performing the first-order Taylor series expansion of term $\mathbf S_k \bm h_{k,s}^{i}$ in \eqref{eq3} around the ${[i-1]}$-th extent $\hat{\bm p}_k^{[i-1]}$ and velocity $\hat{\bm {\vartheta}}_k^{[i-1]}$, respectively, and keeping $\bm h_{k,s}^{[i]}$ as a random variable yields

	\begin{equation} \label{eq47}
		\begin{split}
			\mathbf S_k \bm h_{k,s}^{i}  \approx & \underbrace{\hat{\mathbf{S}}_k^{[i-1]} \boldsymbol{h}_{k,s}^{i}}_{\mathrm{I}}+\underbrace{\left[\begin{array}{c}
					\left(\boldsymbol{h}_{k,s}^{i}\right)^{\mathsf{T}} \hat{\mathbf{J}}_{1k,p}^{[i-1]} \\
					\left(\boldsymbol{h}_{k,s}^{i}\right)^{\mathsf{T}} \hat{\mathbf{J}}_{2k,p}^{[i-1]}
				\end{array}\right]\left(\bm{p}_k-\hat{\bm{p}}_k^{[i-1]}\right)}_{\mathrm{II}} \\
			& + 
			\underbrace{\left[\begin{array}{c}
					\left(\boldsymbol{h}_{k,s}^{i}\right)^{\mathsf{T}} \hat{\mathbf{J}}_{1k,v}^{[i-1]} \\
					\left(\boldsymbol{h}_{k,s}^{i}\right)^{\mathsf{T}} \hat{\mathbf{J}}_{2k,v}^{[i-1]}
				\end{array}\right]\left(\bm{\vartheta}_k-\hat{\bm{\vartheta}}_k^{[i-1]}\right)}_{\mathrm{III}} 
		\end{split}
	\end{equation}
	where $\mathbf{J}_{1k,v}$ and $\mathbf{J}_{2k,v}$ are the Jacobian matrices of the first
	row $\mathbf S_{1,k}$ and second row $\mathbf S_{2,k}$ of $\mathbf S_k$ at the ${[i-1]}$-th velocity estimate $\hat{\bm {\vartheta}}_k^{[i-1]}$ (see \eqref{eq48} and \eqref{eq49}), and $\mathbf{J}_{1k,p}$ and $\mathbf{J}_{2k,p}$ are the Jacobian matrices of the first row $\mathbf S_{1,k}$ and second row $\mathbf S_{2,k}$ of $\mathbf S_k$ at the ${[i-1]}$-th extent estimate $\hat{\bm p}_k^{[i-1]}$ (see \eqref{eq50} and \eqref{eq51}). 
\begin{figure*}[!t]
		\normalsize
		\setcounter{MYtempeqncnt}{\value{equation}}
		\setcounter{equation}{47}
	\begin{equation} \label{eq48}
	\mathbf{J}_{1k,v}  = \left.  \frac{\partial \mathbf S_{1,k}}{\partial \bm {\vartheta}_k} \right|_{\hat{\bm {\vartheta}}_k^{[i-1]}} = \left. \begin{bmatrix}
		l_{k,1} (d3\sin{\beta_k}  +  d1\cos{\beta_k} ) &  l_{k,1} (d2\sin{\beta_k}  + d3\cos{\beta_k} )   \\
		l_{k,2} (d1\sin{\beta_k}  - d3\cos{\beta_k} ) &  l_{k,2} (d3\sin{\beta_k}  - d2\cos{\beta_k} )
	\end{bmatrix} \right|_{\hat{\bm \vartheta}_k^{[i-1]}}, 
	\end{equation}
	\begin{equation} \label{eq49}
		\mathbf{J}_{2k,v}  = \left.  \frac{\textcolor{red} {\partial \mathbf S_{2,k}}}{\partial \bm {\vartheta}_k} \right|_{\hat{\bm {\vartheta}}_k^{[i-1]}} = \left. \begin{bmatrix}
			l_{k,1} (d3\cos{\beta_k}  - d1\sin{\beta_k} ) &  l_{k,1} (d2\cos{\beta_k}  - d3\sin{\beta_k} )   \\
			\textcolor{red}{l_{k,2} (d3\sin{\beta_k}  +  d1\cos{\beta_k} )}  &  l_{k,2} (d2\sin{\beta_k}  + d3\cos{\beta_k} )
		\end{bmatrix} \right|_{\hat{\bm \vartheta}_k^{[i-1]}},  
	\end{equation}

	\begin{equation} \label{eq50}
	\mathbf{J}_{1k,p}  = \left. \frac{\textcolor{red} {\partial \mathbf S_{1,k}}}{\partial \bm p_k} \right|_{\hat{\bm p}_k^{[i-1]}} = \left. \begin{bmatrix}
		\frac{v_k^{\mathsf{x}} \cos{\beta_k} + v_k^{\mathsf{y}} \sin{\beta_k}}{\Vert \bm {\vartheta}_k \Vert} & 0 & \frac{l_{k,1}(v_k^{\mathsf{y}} \cos{\beta_k} - v_k^{\mathsf{x}} \sin{\beta_k})}{\Vert \bm {\vartheta}_k \Vert} \\
		0 & \frac{v_k^{\mathsf{x}} \sin{\beta_k} - v_k^{\mathsf{y}} \cos{\beta_k}}{\Vert \bm {\vartheta}_k \Vert} & \frac{ \textcolor{red}{l_{k,2}}(v_k^{\mathsf{x}} \cos{\beta_k} + v_k^{\mathsf{y}} \sin{\beta_k})}{\Vert \bm {\vartheta}_k \Vert}
	\end{bmatrix} \right|_{\hat{\bm p}_k^{[i-1]}},
	\end{equation}
	\begin{equation} \label{eq51}
		\textcolor{red}{\mathbf{J}_{2k,p}}  = \left. \frac{\partial \mathbf S_{2,k}}{\partial \bm p_k} \right|_{\hat{\bm p}_k^{[i-1]}} = \left. \begin{bmatrix}
			\frac{v_k^{\mathsf{y}} \cos{\beta_k} - v_k^{\mathsf{x}} \sin{\beta_k}}{\Vert \bm {\vartheta}_k \Vert} & 0 & - \frac{l_{k,1}(v_k^{\mathsf{y}} \sin{\beta_k} + v_k^{\mathsf{x}} \cos{\beta_k})}{\Vert \bm {\vartheta}_k \Vert} \\
			0 & \frac{v_k^{\mathsf{y}} \sin{\beta_k} + v_k^{\mathsf{x}} \cos{\beta_k}}{\Vert \bm {\vartheta}_k \Vert} & \frac{\textcolor{red}{l_{k,2}}(v_k^{\mathsf{y}} \cos{\beta_k} - v_k^{\mathsf{x}} \sin{\beta_k})}{\Vert \bm {\vartheta}_k \Vert}
		\end{bmatrix} \right|_{\hat{\bm p}_k^{[i-1]}}.
	\end{equation}
		\setcounter{equation}{\value{MYtempeqncnt}}
		\hrulefill
		\vspace*{2pt}
\end{figure*}
\setcounter{equation}{51}

Substituting \eqref{eq47} into \eqref{eq3}, using the fact that the terms ${\mathrm{II}} $ and ${\mathrm{III}}$ in \eqref{eq47} are scalar, the residual covariance about $\bm y_{k,s}^{i}$ in \eqref{eq3} is calculated as
	\begin{equation} \label{eq52}
		\mathbf C_{k,s}^{y[i]} = \mathbf H \mathbf C_k^{x[i-1]} \mathbf H^{\mathsf{T}} + \mathbf C^{\mathrm{\Rmnnum{1}}} + \mathbf C^{\mathrm{\Rmnnum{2}}} + \mathbf C^{\mathrm{\Rmnnum{3}}} + \mathbf C_s^v,
	\end{equation}
	and then \eqref{eq3} is rewritten as \eqref{eq6}.
	The proof is complete.
\end{IEEEproof}

\section{Proof of proposition \ref{the2}} \label{proof2}

\begin{figure*}[!t]
	\begin{IEEEproof} 
		It is shown that \eqref{eq10} does not exist a directly linear/nonlinear mapping between $\bm p_k$ and $\mathbf Y_{k,s}^{[i]}$. 
		\normalsize
		\setcounter{MYtempeqncnt}{\value{equation}}
		To extract the term $\bm p_k$ from \eqref{eq10}, substituting \eqref{eq47} into \eqref{eq3} yields	
		\setcounter{equation}{52}
		
		
		\begin{equation} \label{eq53}
			\bm y_{k,s}^{i} \approx \; \mathbf H \bm x_k + \hat{\mathbf{S}}_k^{[i-1]} \bm{h}_{k,s}^{i} + \left[\begin{array}{c}
				\left(\boldsymbol{h}_{k,s}^{i}\right)^{\mathsf{T}} \hat{\mathbf{J}}_{1k,p}^{[i-1]} \\
				\left(\boldsymbol{h}_{k,s}^{i}\right)^{\mathsf{T}} \hat{\mathbf{J}}_{2k,p}^{[i-1]}
			\end{array}\right]\left(\bm{p}_k-\hat{\bm{p}}_k^{[i-1]}\right) + \left[\begin{array}{c}
				\left(\boldsymbol{h}_{k,s}^{i}\right)^{\mathsf{T}} \hat{\mathbf{J}}_{1k,v}^{[i-1]} \\
				\left(\boldsymbol{h}_{k,s}^{i}\right)^{\mathsf{T}} \hat{\mathbf{J}}_{2k,v}^{[i-1]}
			\end{array}\right]\left(\bm{\vartheta}_k-\hat{\bm{\vartheta}}_k^{[i-1]}\right) +
			\bm  v_{k,s}^{i} := 
			\tilde{\bm y}_{k,s}^{i}.
		\end{equation}
		Further, substituting \eqref{eq53} into \eqref{eq10} gives
		\begin{subequations} \label{eq54}
			\begin{align}
				\mathbf Y_{k,s}^{[i]} \approx &{} \mathbf F \left( \left( \tilde{\bm y}_{k,s}^{i} - \mathbf H \hat{\bm x}_k^{[i-1]}  \right) \otimes \Big(\cdot\Big) \right) \label{eq54a} \\
				:= &{} [Y_1^2\: Y_2^2 \: Y_1Y_2]^{\mathsf{T}} \label{eq54b}
			\end{align}
		\end{subequations}
		where the notation $\left(\cdot\right)$ denotes the term right before it. 
		
		Next, our goal is using \eqref{eq54} to construct an equivalent measurement model $\mathbf Y_{k,s}^{[i]} \approx \mathbf H^p_k \bm p_k + \bm v_{k,s}^{p[i]}$ about $\bm p_k$, where $\mathbf H^p_k$ is the measurement matrix, and $\bm v_{k,s}^{p[i]}$ is measurement noise. Meanwhile, the first and second moments of the model should equal to the expectation and covariance of \eqref{eq10} as much as possible. For the clarity, denote  
		$\mathbf H \bm x_k - \mathbf H \hat{\bm x}_k^{[i-1]} := [\tilde{x}_{1}\;\tilde{x}_{2}]^{\mathsf{T}}$, $\mathbf H \mathbf C_k^{x[i-1]} \mathbf H^{\mathsf{T}} := \left[\begin{smallmatrix}
			c^x_{11} & c^x_{12} \\ c^x_{21} & c^x_{22}
		\end{smallmatrix}\right] $, $\mathbf C_{k,s}^{y[i]} :=\left[\begin{smallmatrix}
			c^y_{11} & c^y_{12} \\ c^y_{21} & c^y_{22}
		\end{smallmatrix}\right]$, $\hat{\mathbf{S}}_k^{[i-1]} := [\hat{\mathbf{S}}_{1}^{\mathsf{T}}\;\hat{\mathbf{S}}_{2}^{\mathsf{T}}]^{\mathsf{T}}$, $\bm v_{k,s}^{[i]}:=[v_{1}\;v_{2}]^{\mathsf{T}}$, $\mathrm{Cov} (\bm v_{k,s}^{[i]}) := \mathrm{diag}\left[\begin{smallmatrix}
			\sigma_1^2 & 0 \\ 0 & \sigma_2^2
		\end{smallmatrix} \right] $,
		and omit the superscript $[\cdot]$, time index $k$, and sensor node index $s$. Then,  \eqref{eq54b} is further expanded as follows 
		\begin{subequations} \label{eq55}
			\begin{equation} \label{eq55a}
				Y_1^2  =  \left( \tilde{x}_{1} + \hat{\mathbf{S}}_{1}\bm{h}+  \bm{h}^{\mathsf{T}} \hat{\mathbf{J}}_{1,p} (\bm{p}-\hat{\bm{p}}) + \bm{h}^{\mathsf{T}} \hat{\mathbf{J}}_{1,v} (\bm{\vartheta}-\hat{\bm{\vartheta}}) + v_{1}  \right)^2 - 2 \hat{\mathbf{S}}_{1} \mathbf C^h \hat{\mathbf{J}}_{1,p} \bm{p} + 2 \hat{\mathbf{S}}_{1} \mathbf C^h \hat{\mathbf{J}}_{1,p} \bm{p},
			\end{equation}
			\begin{equation} \label{eq55b}
				Y_1^2  =  \left( \tilde{x}_{2} + \hat{\mathbf{S}}_{2}\bm{h}+  \bm{h}^{\mathsf{T}} \hat{\mathbf{J}}_{2,p} (\bm{p}-\hat{\bm{p}})  + \bm{h}^{\mathsf{T}} \hat{\mathbf{J}}_{2,v} (\bm{\vartheta}-\hat{\bm{\vartheta}}) + v_{2} \right)^2  - 2 \hat{\mathbf{S}}_{2} \mathbf C^h \hat{\mathbf{J}}_{2,p} \bm{p}  +  2 \hat{\mathbf{S}}_{2} \mathbf C^h \hat{\mathbf{J}}_{2,p} \bm{p},
			\end{equation}	
			\begin{equation} \label{eq55c}
				\begin{split}
					Y_1Y_2  = & \left( \tilde{x}_{2} + \hat{\mathbf{S}}_{2}\bm{h}+  \bm{h}^{\mathsf{T}} \hat{\mathbf{J}}_{2,p} (\bm{p}-\hat{\bm{p}})  + \bm{h}^{\mathsf{T}} \hat{\mathbf{J}}_{2,v} (\bm{\vartheta}-\hat{\bm{\vartheta}}) + v_{2} \right) \left( \tilde{x}_{1} + \hat{\mathbf{S}}_{1}\bm{h}+  \bm{h}^{\mathsf{T}} \hat{\mathbf{J}}_{1,p} (\bm{p}-\hat{\bm{p}}) + \bm{h}^{\mathsf{T}} \hat{\mathbf{J}}_{1,v} (\bm{\vartheta}-\hat{\bm{\vartheta}}) + v_{1}  \right) \\
					& + \hat{\mathbf{S}}_{2} \mathbf C^h \hat{\mathbf{J}}_{1,p} \bm{p} +
					\hat{\mathbf{S}}_{1} \mathbf C^h \hat{\mathbf{J}}_{2,p} \bm{p}
					- \hat{\mathbf{S}}_{2} \mathbf C^h \hat{\mathbf{J}}_{1,p} \bm{p} -
					\hat{\mathbf{S}}_{1} \mathbf C^h \hat{\mathbf{J}}_{2,p} \bm{p}.
				\end{split}
			\end{equation}			
		\end{subequations}
		
		The expectation of \eqref{eq55} is given as follows:
		\begin{subequations} \label{eq56}
			\begin{equation} \label{56a}
				\mathbb{E} (Y_1^2) =  c_{11}^x + \hat{\mathbf{S}}_{1} \mathbf C^h \hat{\mathbf{S}}_{1}^{\mathsf{T}} + \mathrm{tr} (\mathbf C^p \hat{\mathbf{J}}_{1,v}^{\mathsf{T}} \mathbf C^h \hat{\mathbf{J}}_{1,v}) + \mathrm{tr} (\mathbf C^p \hat{\mathbf{J}}_{1,p}^{\mathsf{T}} \mathbf C^h \hat{\mathbf{J}}_{1,p}) + \sigma_1^2 - 2 \hat{\mathbf{S}}_{1} \mathbf C^h \hat{\mathbf{J}}_{1,p} \bm{p} +  2 \hat{\mathbf{S}}_{1} \mathbf C^h \hat{\mathbf{J}}_{1,p} \bm{p},
			\end{equation}
			\begin{equation} \label{56b}
				\mathbb{E} (Y_2^2) =  c_{22}^x + \hat{\mathbf{S}}_{2} \mathbf C^h \hat{\mathbf{S}}_{2}^{\mathsf{T}} + \mathrm{tr} (\mathbf C^p \hat{\mathbf{J}}_{2,v}^{\mathsf{T}} \mathbf C^h \hat{\mathbf{J}}_{2,v}) + \mathrm{tr} (\mathbf C^p \hat{\mathbf{J}}_{2,p}^{\mathsf{T}} \mathbf C^h \hat{\mathbf{J}}_{2,p}) + \sigma_2^2  - 2 \hat{\mathbf{S}}_{2} \mathbf C^h \hat{\mathbf{J}}_{2,p} \bm{p} +  2 \hat{\mathbf{S}}_{2} \mathbf C^h \hat{\mathbf{J}}_{2,p} \bm{p},
			\end{equation}	
			\begin{equation} \label{56c}
				\mathbb{E} (Y_1 Y_2) =  c_{12}^x + \hat{\mathbf{S}}_{1} \mathbf C^h \hat{\mathbf{S}}_{2}^{\mathsf{T}} + \mathrm{tr} (\mathbf C^p \hat{\mathbf{J}}_{1,v}^{\mathsf{T}} \mathbf C^h \hat{\mathbf{J}}_{2,v}) + \mathrm{tr} (\mathbf C^p \hat{\mathbf{J}}_{1,p}^{\mathsf{T}} \mathbf C^h \hat{\mathbf{J}}_{2,p}) + \hat{\mathbf{S}}_{2} \mathbf C^h \hat{\mathbf{J}}_{1,p} \bm{p} + \hat{\mathbf{S}}_{1} \mathbf C^h \hat{\mathbf{J}}_{2,p} \bm{p}
				- \hat{\mathbf{S}}_{2} \mathbf C^h \hat{\mathbf{J}}_{1,p} \bm{p} -
				\hat{\mathbf{S}}_{1} \mathbf C^h \hat{\mathbf{J}}_{2,p} \bm{p}.
			\end{equation}	
		\end{subequations}
		Equation \eqref{eq54a} implies $\mathbb{E} (Y_1) = \mathbb{E} (Y_2) =0$, $\mathrm{Cov} (Y_1) = c_{11}^y$, $\mathrm{Cov} (Y_1Y_2) = c_{12}^y$ and $\mathrm{Cov} (Y_2) = c_{22}^y$. From Wick's theorem \cite{b23}, we have 
		\begin{equation} \label{eq57}
			\left \{ \begin{lgathered}
				\mathbb{E} \{(Y_1^2)^2\} = 3 (c_{11}^y)^2, \; \mathbb{E} \{(Y_2^2)^2\} = 3 (c_{22}^y)^2, \; \mathbb{E} \{(Y_1^3Y_2)\} = 3 (c_{11}^y c_{12}^y) \\ \mathbb{E} \{(Y_1 Y_2^3)\} = 3 (c_{22}^y c_{12}^y), \; \mathbb{E} \{(Y_1^2 Y_2^2)\} =  c_{11}^y c_{22}^y + 2 (c_{12}^y)^2
			\end{lgathered} \right..
		\end{equation}
		By rearranging \eqref{eq54b}, \eqref{eq55}, \eqref{eq56} and \eqref{eq57}, we have 
		\begin{equation} \label{eq58} 
			\left[\begin{array}{c} Y_1^2  \\ Y_2^2 \\ Y_1 Y_2 \end{array}\right] = \hat{\mathbf M} \bm p + \bm v^p 
		\end{equation}
		with expectation $\mathbf F \mathrm{vect} \left(\mathbf C^{y}\right)$ and covariance $\mathbf F \left( \mathbf C^{y}\otimes \mathbf C^{y} \right) (\mathbf F + \tilde{\mathbf F})^{\mathsf{T}}$. Then, we get the measurement model as shown in \eqref{eq14}, and find the measurement matrix $\mathbf H^p = \hat{\mathbf M}$. The first and second moments of \eqref{eq58} are consistent with the results as shown in \eqref{eq11} and \eqref{eq12}, respectively, which means that \eqref{eq58} matches the moments information in \eqref{eq10}. Notice that using those terms in $\hat{\mathbf M}$ guarantees \eqref{eq54b} and \eqref{eq10} giving the same value on computing the cross-covariance between $\mathbf Y$ and $\bm p$, but choosing other terms do not satisfy this condition. 	
		\setcounter{equation}{\value{MYtempeqncnt}}
		Until now, the proof is complete.
	\end{IEEEproof}	
	\hrulefill
\end{figure*}

\ifCLASSOPTIONcaptionsoff
  \newpage
\fi



%

\begin{IEEEbiography}
	[{\includegraphics[width=1in,height=1.25in,clip,keepaspectratio]{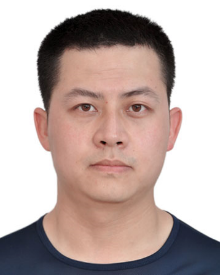}}]{Zhifei Li}  received the M.E. and Ph.D. degrees in Information and Communication Engineering from the National University of Defense Technology, Hefei, China, in 2016 and 2020, respectively. During September 2019--July 2020, he was a Visiting Research Scalar in the the School of Automation at Northwestern Polytechnical University, Xi’an, China. 
	He is currently a Lecturer with the School of Space Information, Space Engineering University, Beijing, China. His research interests include statistical signal processing, extended object tracking, multi-sensor control, distributed system design and data fusion.
\end{IEEEbiography}

\begin{IEEEbiography}
	[{\includegraphics[width=1in,height=1.25in,clip,keepaspectratio]{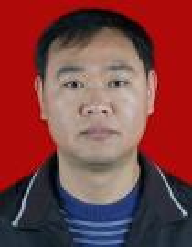}}%
	]{Yan Liang} received the B.E., M.E., and Ph.D. degrees from Northwestern Polytechnical University (NPU), Xi’an, China, in 1993, 1998, and 2001, respectively. He was a Postdoctoral Fellow with Tsinghua University, Beijing, China, from 2001 to 2003, a Research Fellow with the Hong Kong Polytechnic University, Hong Kong, from 2006 to 2006, and a Visiting Scholar with the University of Alberta, Edmonton, AB, Canada, from 2007 to 2008. Since 2009, he has been a Professor with the School of Automation, NPU. His research interests include estimation theory, information fusion, and target tracking.
\end{IEEEbiography}
	
\begin{IEEEbiography}
	[{\includegraphics[width=1in,height=1.25in,clip,keepaspectratio]{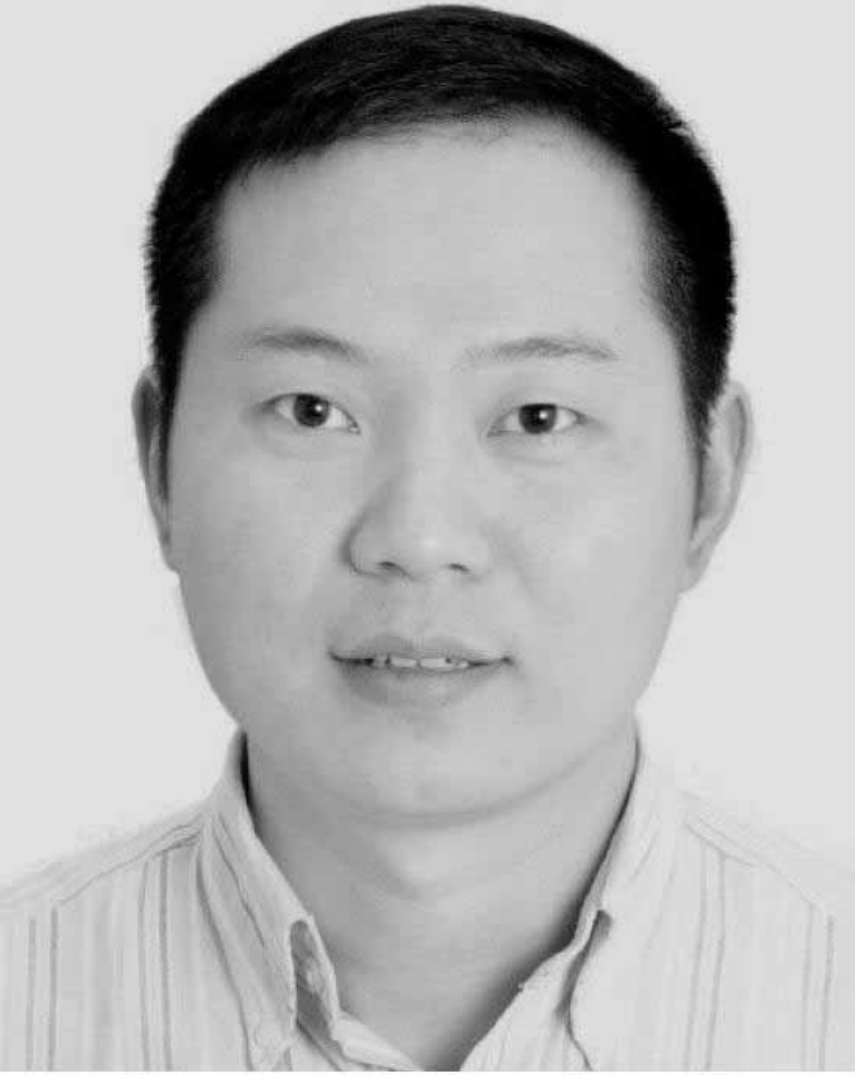}}%
	]{Linfeng Xu} (S'10-M'14) received the B.S. and M.S. degrees in mechatronic
	engineering from Northwestern Polytechnical University (NPU), Xi'an, China, in 2002 and 2005, respectively, and the Ph.D. degree in Control Science and	Engineering from Xi'an Jiaotong University, China, in 2013.
	
	In October 2013, he joined the School of Automation at NPU and is currently
	working as an associate professor. During September 2017--September 2018, he was a visiting research fellow in the Department of Electrical Engineering at the University of New Orleans, LA, USA. He is also a member of the Laboratory of Information Fusion Technology, Ministry of Education, China. His research interests are in detection and estimation theory, target tracking, and air traffic control.
\end{IEEEbiography}

%

%
%
%




\end{document}